\documentclass{ptephy_v1}

\newtheorem{theorem}{Theorem}
\newtheorem{definition}{Definition}
\newtheorem{proposition}{Proposition}

\begin{document}

\title{Extension of photon surfaces and their area:
  Static and stationary spacetimes}



\author{Hirotaka Yoshino${}^1$}
\author{Keisuke Izumi${}^{2,3}$}
\author{Tetsuya Shiromizu${}^{3,2}$}
\author{Yoshimune Tomikawa${}^4$}
\affil{${}^1$Department of Mathematics and Physics, Osaka City University, Osaka 558-8585, Japan}
\affil{${}^2$Kobayashi-Maskawa Institute, Nagoya University, Nagoya 464-8602, Japan}
\affil{${}^3$Department of Mathematics, Nagoya University, Nagoya 464-8602, Japan}
\affil{${}^4$Faculty of Economics, Matsuyama University, Matsuyama 790-8578, Japan}


\begin{abstract}
  We propose a new concept, the transversely trapping surface (TTS),
  as an extension of the static photon surface
  characterizing the strong gravity region 
  of a static/stationary spacetime in terms of photon behavior.
  The TTS is defined as a static/stationary timelike surface $S$ 
  whose spatial section is a 
  closed two-surface, such that
  arbitrary photons emitted tangentially to $S$
  from arbitrary points on $S$ propagate on or toward the inside of $S$.
  We study the properties of TTSs 
  for static spacetimes and axisymmetric stationary spacetimes.
  In particular, the area $A_0$ of a TTS is proved to be bounded as
  $A_0\le 4\pi(3GM)^2$ under certain conditions, where $G$
  is the Newton constant and $M$ is the total mass. 
  The connection between the TTS and the loosely trapped
  surface proposed by us [arXiv:1701.00564] is also examined. 
\end{abstract}

\subjectindex{E0, E31, A13}

\maketitle

%
\section{Introduction}
\label{section1}

If a black hole forms, everything is trapped inside of
its horizon. Such extremely strong gravity is
realized only when a mass is concentrated in a small region.
One of the mathematical conjectures concerning its scale
is the Penrose inequality \cite{Penrose:1973},
\begin{equation}
  A_{\rm H}\le 4\pi (2GM)^2,
  \label{Penrose-inequality}
\end{equation}
where $A_{\rm H}$ is the area of an apparent horizon, $G$ is 
the Newton gravitational constant, and $M$ is the
Arnowitt--Deser--Misner (ADM) mass. Here, the right-hand side is
the horizon area of the Schwarzschild black hole 
with the same mass. The Penrose inequality
has been proved with the methods of
the inverse mean curvature flow \cite{Wald:1977,Huisken:2001}
and the conformal flow \cite{Bray:2001}
for time-symmetric initial data with nonnegative Ricci scalar.

In a Schwarzschild spacetime, a collection of 
unstable circular orbits of null geodesics
forms a sphere at $r=3GM$, called a photon sphere.
The photon sphere plays an
important role in phenomena related to observations, like 
gravitational lensing \cite{Virbhadra:1999}
and the ringdown of waves around a
black hole \cite{Cardoso:2016}. The region between
the event horizon and the photon sphere is a very characteristic 
region because if photons are emitted isotropically from a point
in this region, more than half of them 
will be (eventually) trapped by the horizon \cite{Synge:1966}
(see also Sect.~5.1 of \cite{Perlick:2004}).

A nonspherical black hole 
would also possess a strong gravity region
in which (roughly speaking) photons propagating in
the transverse direction to the source will be trapped
by the horizon. 
It is a basic problem to determine/constrain
its characteristic scale. To be more specific, 
we expect that if an appropriate definition is given, 
an inequality that is analogous to the Penrose inequality
should hold for surfaces in such a strong gravity region,
that is, 
\begin{equation}
  A_0\le 4\pi (3GM)^2.
  \label{Penroselike-inequality}
\end{equation}
Here, $A_0$ is the area of a surface in the strong gravity
region and the right-hand side is
the area of the photon sphere of a Schwarzschild spacetime 
with the same mass.
In this paper, we call this inequality {\it the Penrose-like inequality}.
In order to formulate and prove the Penrose-like inequality,
we have to introduce an appropriate concept
of a surface characterizing the strong gravity region.

One of the generalized concepts of the photon sphere
is the photon surface \cite{Claudel:2000}. It is defined
as a timelike hypersurface $S$ such that any photon emitted
in any tangent direction of $S$ from any point on $S$
continues to propagate on $S$. The photon surface is allowed to be dynamical
or to be non spherically symmetric.
In our context, the concept of the static photon surface 
may be expected to be useful to characterize the strong gravity region.
However, the existence of a photon surface practically requires
high symmetry of the spacetime, because the condition of a photon surface 
strongly constrains the photon behavior on it.
For this reason, the uniqueness of static photon surfaces has been expected
and partially proved \cite{Cederbaum:2014,Cederbaum:2015a,Yazadjiev:2015a,Cederbaum:2015b,Yazadjiev:2015b,Yazadjiev:2015c,Rogatko:2016,Yoshino:2016,Tomikawa:2016,Tomikawa:2017};
if a static photon surface exists, the
spacetime must be spherically symmetric in various setups
(see also an example of a nonspherical photon surface
but with a conical singularity \cite{Gibbons:2016}).
Another manifestation of the strong requirement on a
photon surface is that it does not exist in stationary spacetimes.
In a Kerr spacetime, for example, 
there are null geodesics staying on $r=\mathrm{const.}$ surfaces,
but their $r$ values depend on the angular momentum of the
photons \cite{Teo:2003}.
As a result, a collection of
photon orbits with constant $r$ values forms a photon
region with thickness 
instead of a photon surface (Sect.~5.8
of \cite{Perlick:2004}).
The photon region becomes infinitely thin
and reduces to a photon surface in the limit of zero rotation.

Clearly, the absence of a photon surface does not imply the
absence of a strong gravity region. 
It is nice to introduce other concepts
to characterize the strong gravity region 
that are applicable to spacetimes without high symmetry
or to stationary spacetimes. 
One such approach is 
{\it the loosely trapped surface (LTS)} proposed by us \cite{Shiromizu:2017}.
An LTS is defined with 
quantities of intrinsic geometry of the initial data;
in a Schwarzschild spacetime, 
the marginal LTS coincides with the photon sphere. 
We have proved that LTSs satisfy
the Penrose-like inequality \eqref{Penroselike-inequality}
in initial data with a nonnegative Ricci scalar. However, 
the connection between the LTS and the photon behavior
in non-Schwarzschild cases is still unclear.
We speculated that such a connection
would exist, but it was left as a remaining problem.

In light of the above discussion, 
the purpose of this paper is threefold.
First, as a generalization of the static photon surface,
we introduce a new concept 
to characterize a strong gravity region through the behavior of photons,
{\it the transversely trapping surface (TTS)}. 
A TTS is defined as a static/stationary timelike hypersurface $S$ such that
arbitrary photons emitted tangentially to $S$ propagate on
or toward the inside of $S$.
TTSs can be present in static spacetimes
without high symmetry and in stationary spacetimes, and 
examples in a Kerr spacetime are explicitly calculated.
Second, we show how the TTS is related
to the LTS in static spacetimes
and axisymmetric stationary spacetimes.
Through this study, we give an answer to the remaining
problem of our previous
paper. 
Third, we will prove that TTSs satisfy the Penrose-like
inequality~\eqref{Penroselike-inequality}
in static spacetimes and axisymmetric stationary spacetimes 
under fairly generic conditions.

This paper is organized as follows.
In the next section, we explain the basic concepts
and properties of the TTS and the LTS.
In Sect.~\ref{section3}, we study TTSs in static spacetimes. 
In Sect.~\ref{section4}, TTSs in axisymmetric stationary spacetimes
are explored. 
Section~\ref{section5} is devoted to a summary and discussion.
In Appendix~A, we present derivations of the TTS condition
in static and axisymmetric-stationary cases
in a different manner from those in the main article.
In Appendix~B, we give a supplementary explanation of the derivation of the
TTS condition in the axisymmetric stationary case.
In Appendix~C, we present some geometric formulas that are useful
for studying TTSs in more general cases.
Throughout the paper, we use units in which $c=1$.
Although we write the Newton constant $G$ basically,
we set $G=1$ when a Kerr spacetime is studied in Sect.~\ref{section4-2}
for conciseness.

%
%
\section{Definitions of surfaces in strong gravity regions}
\label{section2}

In this section, we explain the two concepts of surfaces
to characterize strong gravity regions. In Sect.~\ref{section2-1},
we define the TTS
and derive the mathematical condition for
a surface $S$ to be a TTS.
In Sect.~\ref{section2-2}, we review the LTS
that was introduced in our previous
paper \cite{Shiromizu:2017}. A theorem proved in \cite{Shiromizu:2017},
which is used in this paper as well, is also reviewed.

\subsection{Transversely trapping surfaces}
\label{section2-1}

Consider a static or stationary spacetime $\mathcal{M}$ possessing
a timelike Killing vector field $t^a$, and take
a spacelike hypersurface $\Sigma$ given by $t=\mathrm{const.}$
We consider an orientable closed
two-surface $S_0$ in $\Sigma$ and suppose that $\Sigma$
is divided into the inside and outside regions by $S_0$.
By transporting $S_0$ along the integral lines of $t^a$,
we obtain a static/stationary three-dimensional timelike surface $S$.
In this setup, we define the TTS as follows:
%
\begin{definition}
  A static/stationary timelike
    hypersurface $S$ is a TTS if and only if arbitrary light rays
    emitted in arbitrary tangential directions
    of $S$ from arbitrary points of $S$ propagate on $S$ or
    toward the inside region of $S$.
\label{definition-1}
\end{definition}
%

%
\begin{figure}[tb]
\centering
\includegraphics[width=0.7\textwidth,bb=0 0 728 373]{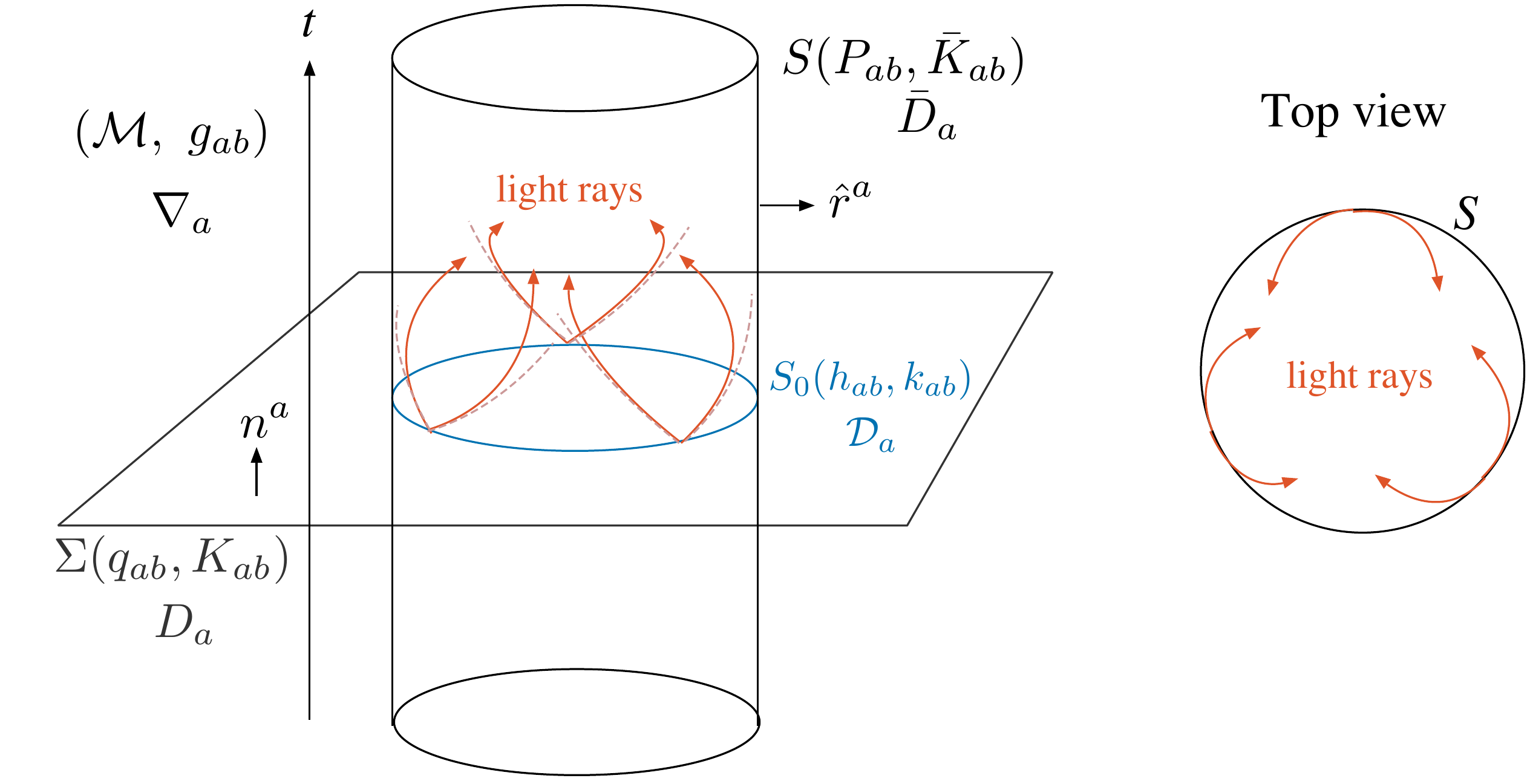}
\caption{
  Schematic picture of the transversely
  trapping surface. The notations are also indicated.
}
\label{schematic-TTS}
\end{figure}
%

Below, we derive the mathematical expression for the condition
for $S$ to be a TTS.
Before starting our analysis, we summarize the notations commonly 
used throughout this paper. 
The metric of the spacetime $\mathcal{M}$ is $g_{ab}$.
With the future-directed unit normal $n^a$
to $\Sigma$, the induced metric 
and the extrinsic curvature of $\Sigma$ are given by 
$q_{ab}=g_{ab}+n_an_b$ and $K_{ab} = (1/2)\pounds_nq_{ab}$,
respectively, where $\pounds$ denotes the
Lie derivative.
With the outward unit normal $\hat{r}^a$ to $S$, 
the induced metric
and the extrinsic curvature of $S$ are 
$P_{ab}=g_{ab}-\hat{r}_a\hat{r}_b$ and
$\bar{K}_{ab}=(1/2)\pounds_{\hat{r}}P_{ab}$, respectively.
Although the unit normal to $S$ in $\mathcal{M}$ and the unit normal to $S_0$ in $\Sigma$
do not coincide in general, in this paper
we consider setups such that these two unit normals
agree. We will come back to this point
in Sects.~\ref{section3-1} and \ref{section4-1}.
In this situation, 
the induced metric of $S_0$ is $h_{ab}=g_{ab}+n_an_b-\hat{r}_a\hat{r}_b$,
and its extrinsic curvature in the hypersurface $\Sigma$ is
$k_{ab}=(1/2){}^{(3)}\pounds_{\hat{r}}h_{ab}$, where ${}^{(3)}\pounds$
is the Lie derivative on the hypersurface $\Sigma$. 
The covariant derivatives of $\mathcal{M}$, $\Sigma$, $S$, and $S_0$
are denoted as $\nabla_a$, $D_a$, $\bar{D}_a$, and $\mathcal{D}_a$,
respectively.
These definitions are summarized in Fig.~\ref{schematic-TTS}.

Consider a null geodesic $\gamma$ with the tangent vector $k^a$ 
emitted tangentially to $S$ from a point $p$ on $S$.
If $S$ is a TTS, 
this null geodesic must stay on $S$ or go toward the inside of
$S$. Let us introduce another null trajectory $\bar{\gamma}$
from the point $p$ with the tangent vector $\bar{k}^a$, which 
is assumed to be a null geodesic on the hypersurface $S$,
\begin{equation}
  \bar{k}^a\bar{D}_a\bar{k}^c=0.
  \label{Geodesic-HypersurfaceS}
\end{equation}
At the point $p$, we choose $\bar{k}^a$
to be same as $k^a$, i.e. $k^a=\bar{k}^a$.
Rewriting Eq.~\eqref{Geodesic-HypersurfaceS}
in terms of the four-dimensional quantities, we have
\begin{equation}
a^c = -(\bar{K}_{ab}\bar{k}^a\bar{k}^b)\hat{r}^c,
\end{equation}
where $a^c$ is four-acceleration, $a^c:=\bar{k}^a\nabla_a\bar{k}^c$.
The two trajectories $\gamma$ and $\bar{\gamma}$
agree locally when $\bar{K}_{ab}\bar{k}^a\bar{k}^b=0$,
and the four-acceleration $\bar{a}^c$ 
of $\bar{\gamma}$ is directed toward the outside 
if and only if $\bar{K}_{ab}\bar{k}^a\bar{k}^b< 0$.
This means that a four-dimensional null geodesic
$\gamma$ satisfies the desired property
if and only if $\bar{K}_{ab}k^ak^b\le 0$ holds.
This result is summarized as follows:
%
\begin{proposition}
  The necessary and sufficient condition for $S$ to be a TTS
  is that for every point on $S$, the condition
\begin{equation}
  \bar{K}_{ab}k^ak^b\le 0
  \label{TTS-condition}
\end{equation}
holds for an arbitrary null tangent vector $k^a$ of $S$.
\label{proposition-1}
\end{proposition}
%

Hereafter, we call this condition {\it the TTS condition}.
In Sects.~\ref{section3-1} and \ref{section4-1},
we will rewrite the TTS condition in the cases of
static spacetimes and axisymmetric stationary spacetimes.
Note that if the equality in \eqref{TTS-condition}
holds at all points on $S$, the surface $S$
coincides with the photon surface
proposed in \cite{Claudel:2000}. Therefore,
our definition of the TTS includes the photon surface
as the marginal case.

In addition to the TTS condition, we sometimes require
the two-surface $S_0$ to be a convex surface.
The condition for the convexity 
depends on the choice of the slice $t=\mathrm{const}.$
We will specify this point in Sects.~\ref{section3-1} and \ref{section4-1}.
When the convex condition is additionally imposed on the TTS,
we call it {\it the convex TTS}.

\subsection{Loosely trapped surfaces}
\label{section2-2}

Now, we explain the LTS defined in our previous paper \cite{Shiromizu:2017}.
Consider the initial data $\Sigma$
of a (not necessarily static or stationary) spacetime $\mathcal{M}$
and a closed two-surface $S_0$ that
divides $\Sigma$ into outside and inside regions.
The initial data $\Sigma$ is supposed to have
a nonnegative Ricci scalar ${}^{(3)}R\ge 0$.
We introduce a radial foliation of $\Sigma$ starting from $S_0$
specified by the coordinate $r$ with the dual basis
$D_ar=\hat{r}_a/\varphi$, 
where $\hat{r}_a$ is the unit normal to $S_0$
and $\varphi$ is the (spatial) lapse function.
In this setup, the LTS is defined as follows:
%
\begin{definition}
  The surface $S_0$ is a loosely trapped surface
  if $k>0$ and $\hat{r}^aD_ak\ge 0$, where $k$ is the trace
  of the extrinsic curvature $k_{ab}$.
\label{definition-2}
\end{definition}
%

The motivation for this definition is that
in the static slice of the Schwarzschild spacetime,
the $r=\mathrm{const.}$ surface has this property
in the range $2M< r\le 3M$. Since $r=3M$ is the
photon sphere, photons are loosely trapped
inside of this sphere and the positivity of $\hat{r}^aD_ak$
is expected to be a useful indicator for a strong gravity region.
Note that, as seen from the following formula which is derived from the
trace of the Ricci equation with the double trace of the Gauss equation,
\begin{equation}
    \hat{r}^aD_ak = -\frac12\left({}^{(3)}R-{}^{(2)}R+k^2+k_{ab}k^{ab}\right)
    -\frac{1}{\varphi}\mathcal{D}^2\varphi,
    \label{LTS-Ricci-equation}
\end{equation}
the value of $\hat{r}^aD_ak$ depends on the
choice of the lapse function $\varphi$.
The surface $S_0$ is called an LTS if $\hat{r}^aD_ak\ge 0$
is satisfied (at least) for one choice of $\varphi$.

There are two main results in our previous paper \cite{Shiromizu:2017}.
The first is that the LTS has topology $S^2$
and satisfies
\begin{equation}
  \int_{S_0}k^2dA \le \frac{16}{3}\pi.
  \label{integral-k2dA}
\end{equation}
This is proved by integrating the relation~\eqref{LTS-Ricci-equation}
under the condition $\hat{r}^aD_ak\ge 0$ and ${}^{(3)}R\ge 0$ and
using the Gauss--Bonnet theorem. The second result is
very important in this paper, and we state it in the form of a theorem:
%
\begin{theorem}
  If the inequality \eqref{integral-k2dA} is satisfied on $S_0$ with
  $k\ge 0$ in asymptotically flat initial data $\Sigma$ with
  nonnegative Ricci scalar ${}^{(3)}R$,
  the area $A_0$ of the surface $S_0$ satisfies the Penrose-like inequality
  \eqref{Penroselike-inequality}, $A_0\le 4\pi (3GM)^2$.
  \label{theorem-1}
\end{theorem}
%

In order to prove this,
we used the method of the inverse mean curvature
flow originally proposed in \cite{Geroch:1973,Wald:1977}.
The inverse mean curvature flow is 
generated by the lapse function $\varphi=1/k$, and along this
flow, Geroch's quasilocal energy is 
monotonic and asymptotes to
the ADM mass at spacelike infinity
$r\to \infty$.\footnote{Because this property
of Geroch's mass is used in the proof, our theorems apply
to asymptotically flat spacetimes. 
Note that the modification of Geroch's mass has been
proposed for asymptotically anti-de Sitter spacetimes \cite{Wang:2001}.}
This leads to the bound on the surface area
(see also \cite{Huisken:2001} for
resolution of the possible formation of singularities along the
flow). We refer to our previous paper \cite{Shiromizu:2017}
for the detailed proof.  
Note that the theorem 
in \cite{Shiromizu:2017} states that  
the Penrose-like inequality holds for an LTS.
We modified the statement of the theorem to the above 
because the inequality \eqref{integral-k2dA}
and $k\ge 0$ on $S_0$ are necessary
in the proof, but the LTS condition $\hat{r}^aD_ak\ge 0$
is not used directly; it was used to guarantee \eqref{integral-k2dA}
in our previous paper.  
This modification will become important in Sects.~\ref{section3-3} and
\ref{section4-4}.

%
%
\section{Static spacetimes}
\label{section3}

In this section, we explore the properties of
TTSs in static spacetimes. In Sect.~\ref{section3-1},
we explain the setup and rewrite 
the TTS condition.
In Sect.~\ref{section3-2}, the relation between
the TTS and the LTS is examined
using the Einstein equations.
The Penrose-like inequality for TTSs
is proved in Sect.~\ref{section3-3}.

\subsection{Setup and TTS condition}
\label{section3-1}

A static spacetime
has the property that the
timelike Killing vector field $t^a$ is hypersurface orthogonal \cite{Wald}.
Namely, there exist $t=\mathrm{const.}$ slices on which
\begin{equation}
t^a = \alpha n^a
\end{equation}
holds (called static slices). Here, $\alpha$ is the lapse function. 
On this slice, the extrinsic curvature vanishes, $K_{ab}=0$.
We will rewrite the TTS condition \eqref{TTS-condition}
using this slice. 
For this reason, our results below cannot be applied to the
$t=\mathrm{const.}$ slice with nonvanishing
$K_{ab}$.
An example of static slices 
in a Schwarzschild spacetime is the $t=\mathrm{const.}$
hypersurfaces in the standard Schwarzschild coordinates $(t,r,\theta,\phi)$,
and a counter-example is those of the Gullstrand--Painlev\'e
coordinates \cite{Gullstrand:1922}.

For a static slice $\Sigma$, the unit normal to $S$ in $\mathcal{M}$
and that to $S_0$ in $\Sigma$ agree, and we denote
those common normals as $\hat{r}^a$.
It is easily derived that 
\begin{equation}
\bar{K}_{ab} = -n_an_b\frac{\hat{r}^cD_c\alpha}{\alpha} + k_{ab}
\end{equation}
holds. Since the null tangent vector of $S$ is expressed as
$k^a = n^a + s^a$, where $s^a$ is a unit tangent vector of $S_0$,
the TTS condition \eqref{TTS-condition} is rewritten as
\begin{equation}
  k_{ab}s^as^b\le \frac{\hat{r}^cD_c\alpha}{\alpha}
  \label{TTS-condition-static-1}
\end{equation}
for an arbitrary unit tangent vector $s^a$ of $S_0$. If we introduce
a tensor
\begin{equation}
  \sigma_{ab}:=k_{ab}-\frac{\hat{r}^cD_c\alpha}{\alpha}h_{ab},
  \label{matrix-sigma}
\end{equation}
the TTS condition \eqref{TTS-condition-static-1} is equivalent
to $\sigma_{ab}$ having two nonpositive eigenvalues.
Such conditions are given by
\begin{equation}
  \mathrm{tr}(\sigma_{ab})\le  0\qquad \textrm{and} \qquad
  \mathrm{det}(\sigma_{ab})\ge  0.
\end{equation}
After calculation, these two conditions
can be expressed in a unified form,
\begin{equation}
  \frac{k}{2}+\sqrt{\frac12\tilde{k}_{ab}\tilde{k}^{ab}}\le \frac{\hat{r}^cD_c\alpha}{\alpha},
  \label{TTS-condition-static-2}
\end{equation}
where $k$ is the trace of $k_{ab}$ and
$\tilde{k}_{ab}$ is the trace-free part of $k_{ab}$, i.e. 
$\tilde{k}_{ab} : =k_{ab}-(k/2)h_{ab}$.
This is the TTS condition in the static case. 
Note that the condition \eqref{TTS-condition-static-2}
in the static case can be also derived by studying the
geodesic equations directly. This is demonstrated in Appendix~\ref{appendix-a1}.

For practical purposes, 
the following form may be more useful.
Since $k_{ab}$ is a symmetric tensor,
it can be diagonalized by appropriately choosing
the tetrad basis $(\mathbf{e}^1)_a$ and $(\mathbf{e}^2)_a$ as
\begin{equation}
  k_{ab} = k_1(\mathbf{e}^1)_a(\mathbf{e}^1)_b+
  k_2(\mathbf{e}^2)_a(\mathbf{e}^2)_b.
  \label{kab-Tetrad-static}
\end{equation}
Without loss of generality, we can assume $k_1\ge k_2$.
Using this form of $k_{ab}$, the TTS condition becomes
\begin{equation}
  k_1\le\frac{\hat{r}^aD_a\alpha}{\alpha}.
  \label{TTS-condition-static-3}
\end{equation}
As defined in Sect.~\ref{section2-1}, we call $S$
a convex TTS when $S_0$ is a convex surface.
The surface $S_0$ is a convex surface if and only if both $k_1$
and $k_2$ are nonnegative. 
Therefore, for $S$ to be a convex TTS,
we require $k_2\ge 0$ in addition to the condition~\eqref{TTS-condition-static-3}.
The convex condition can also be expressed in a covariant manner
as $k\ge \sqrt{2\tilde{k}_{ab}\tilde{k}^{ab}}$.

\subsection{Connection to the LTSs}
\label{section3-2}

Below, we study the relation between convex TTSs
and the LTSs. Specifically, the condition
that a convex TTS becomes an LTS simultaneously is investigated
using the Einstein equations.
The projected components of the energy-momentum tensor $T_{ab}$
are defined as 
\begin{equation}
  \rho : =T_{ab}n^an^b, \qquad
  J_a : =-T_{bc}n^b{q^{c}}_{a},\qquad
  S_{ab} : =T_{cd}{q^{c}}_{a}{q^{d}}_{b}.
  \label{Energy-Momentum-Tensor-Projection}
\end{equation} 
We adopt the $3+1$ split form of the Einstein equations.
For a static spacetime, 
\begin{subequations}
\begin{eqnarray}
  {}^{(3)}R&=&16\pi G\rho,
  \label{Einstein-Static-1}\\
  {}^{(3)}R_{ab} &=&\frac{1}{\alpha}D_aD_b\alpha+8\pi G
  \left[S_{ab}+\frac12 q_{ab}(\rho-{S^c}_{c})\right].
  \label{Einstein-Static-2}
\end{eqnarray}
\end{subequations}
Here, Eq.~\eqref{Einstein-Static-1} is the Hamiltonian constraint,
and Eq.~\eqref{Einstein-Static-2} is
the evolution equation with $\pounds_tK_{ab}=0$. 
The momentum constraint is trivially satisfied with $J_a=0$.
Taking the trace of Eq.~\eqref{Einstein-Static-2}, we have
\begin{equation}
  \frac{1}{\alpha}D^2\alpha=4\pi G\left(\rho+{S^c}_{c}\right).
  \label{Einstein-Static-3}
\end{equation}

Consider a convex TTS whose spatial section is $S_0$.
The quantity $\hat{r}^aD_ak$ for $S_0$
is given by the formula~\eqref{LTS-Ricci-equation}.
The three-dimensional Ricci scalar ${}^{(3)}R$ appearing in this equation can be expressed by
the Gauss equation for $S_0$,
\begin{equation}
  {}^{(3)}R =2{}^{(3)}R_{ab} \hat{r}^a\hat{r}^b
  +{}^{(2)}R-k^2+k_{ab}k^{ab}.
  \label{Gauss-SpacelikeHypersurface}
\end{equation}
The first term on the right-hand side of Eq.~\eqref{Gauss-SpacelikeHypersurface} can be written as
\begin{equation}
  {}^{(3)}R_{ab}  \hat{r}^a\hat{r}^b=   8\pi G(\rho+P_r)   
    -\frac{1}{\alpha}\mathcal{D}^2\alpha
    -k\frac{\hat{r}^cD_c\alpha}{\alpha},
    \label{Static-Einstein-rr}
\end{equation}
which is obtained by multiplying 
Eq.~\eqref{Einstein-Static-2} by $\hat{r}^a\hat{r}^b$ and rewriting with
Eq.~\eqref{Einstein-Static-3}.
Here, we introduced pressure in the radial direction,
\begin{equation}
  P_r : =T_{ab}\hat{r}^a\hat{r}^b.
  \label{Radial-Pressure}
\end{equation}
As a result, the formula~\eqref{LTS-Ricci-equation} is transformed into
(see also Appendix~C)
\begin{equation}
  \hat{r}^aD_ak
  =
  -8\pi G(\rho+P_r) 
    +\frac{1}{\alpha}\mathcal{D}^2\alpha
    +k\frac{\hat{r}^cD_c\alpha}{\alpha}
    -k_{ab}k^{ab}-\frac{1}{\varphi}\mathcal{D}^2\varphi. \label{rDtrk}
\end{equation}
For this formula, we consider the condition
that $S_0$ is guaranteed to satisfy the LTS condition $\hat{r}^aD_ak\ge 0$.
Because the lapse function $\varphi$ can be freely chosen,
we impose $\varphi=\alpha$ (if one considers the Schwarzschild
spacetime, this corresponds to adopting the tortoise coordinate for $r$).
The null energy condition
indicates $\rho+P_r\ge 0$ in general, and hence we restrict to the situation
$\rho+P_r=0$ on $S_0$.
Using the expression \eqref{kab-Tetrad-static}
for $k_{ab}$ and applying
the TTS condition \eqref{TTS-condition-static-3}
with the convex condition $k_2\ge 0$, we have
\begin{equation}
k\frac{\hat{r}^cD_c\alpha}{\alpha}
    -k_{ab}k^{ab} \ge k_2(k_1-k_2) \ge 0.
\end{equation}
Therefore, we have found the following:
%
\begin{proposition}
A convex TTS $S$ in a static spacetime
is an LTS as well if $\rho+P_r=0$ on $S$.
\label{proposition-2}
\end{proposition}
%

Note that the condition $\rho+P_r=0$ is not too strong
because it must be imposed just on $S$ and it is
satisfied if the region around $S$ is vacuum.
Also, the Reissner--Nordstr\"om spacetimes
satisfy this condition for $r=\mathrm{const.}$ surfaces.
In this sense, we have proved the close connection
between the TTS and the LTS with sufficient generality.

\subsection{Area bound for TTSs}
\label{section3-3}

Once a TTS is proved to be an LTS, it possesses the properties
that have been proved for LTSs. In particular, its area
satisfies the Penrose-like inequality~\eqref{Penroselike-inequality},
$A_0\le 4\pi (3GM)^2$. 
However, there might be the case
that a TTS is not guaranteed to be
an LTS but satisfies the Penrose-like inequality
(i.e., we suppose that $\rho+P_{r}=0$ may not be necessary on $S$).
Therefore, 
there remains a possibility that the condition
can be relaxed if just the area bound is considered.
Let us explore this possibility.

The strategy here is to find the condition
that a convex TTS $S$ satisfies the inequality \eqref{integral-k2dA}
without using the concept of the LTS, because 
the Penrose-like inequality follows
from \eqref{integral-k2dA}
by Theorem \ref{theorem-1} in Sect.~\ref{section2-2}.
Eliminating ${}^{(3)}R$ and $\hat{r}^a\hat{r}^bR_{ab}$
from Eqs.~\eqref{Einstein-Static-1}, \eqref{Gauss-SpacelikeHypersurface},
and \eqref{Static-Einstein-rr}, we have (see also Appendix C)
\begin{equation}
  {}^{(2)}R = -16\pi GP_r
  +\frac{2}{\alpha}{\mathcal{D}^2\alpha}
  +2k\frac{\hat{r}^aD_a\alpha}{\alpha}
  +k^2-k_{ab}k^{ab}.
  \label{2D-RicciScalar-static}
\end{equation}
Using the expression \eqref{kab-Tetrad-static} for $k_{ab}$
and the TTS condition \eqref{TTS-condition-static-3},
we have
\begin{eqnarray}
2k\frac{\hat{r}^aD_a\alpha}{\alpha}
+k^2-k_{ab}k^{ab}
\ge 
\frac{3}{2}k^2 + \frac12(k_1+3k_2)(k_1-k_2)
\ge 
\frac{3}{2}k^2,
\label{Inequality-for-theorem-Static}
\end{eqnarray}
for a convex TTS. Assuming $P_r\le 0$ and 
integrating
the relation \eqref{2D-RicciScalar-static} over $S_0$,
we have
\begin{equation}
  \int_{S_0}{}^{(2)}RdA
  \ge
  \int_{S_0}
  \left(  \frac{3}{2}k^2
  +\frac{2\mathcal{D}_a\alpha\mathcal{D}^a\alpha}{\alpha^2}
  \right)dA.
\end{equation}
If $k>0$ at least
at one point, 
the Gauss--Bonnet theorem tells us
that $S_0$ has topology $S^2$ and
the left-hand side is $\int{}^{(2)}RdA=8\pi$. This 
implies the inequality \eqref{integral-k2dA}.
Therefore, we have found the following:
%
\begin{theorem}
  The static time cross section of a convex TTS, $S_0$,
  has topology $S^2$ and  
  satisfies the Penrose-like inequality
  $A_0\le 4\pi (3GM)^2$ if $P_r\le 0$ holds on $S_0$,
  $k>0$ at least at one point on $S_0$, and ${}^{(3)}R$
  is nonnegative (i.e. the energy density $\rho\ge 0$)
  in the outside region on a static slice in an asymptotically
  flat static spacetime.
  \label{theorem-2}
\end{theorem}
%

Compared to Proposition \ref{proposition-2} in Sect.~\ref{section3-2},
the equality $\rho+P_r=0$ is relaxed to the
inequality $P_r\le 0$. Therefore, Theorem \ref{theorem-2}
is expected to have greater applicability.
Note that from Eq.~\eqref{Inequality-for-theorem-Static},
this theorem also holds for a nonconvex TTS if $k_2$ is within the
range $k_2\ge -k_1/3$.

%
%
\section{Axisymmetric stationary spacetimes}
\label{section4}

In this section, we explore the properties of
axisymmetric TTSs in (nonstatic) stationary spacetimes with
axial symmetry. In Sect.~\ref{section4-1},
we explain the setup in detail and
rewrite the TTS condition.
In Sect.~\ref{section4-2}, we show that TTSs actually
can exist in a stationary spacetime by 
presenting examples of a Kerr spacetime. 
In Sect.~\ref{section4-3}, the relation between
the TTS and the LTS is examined, but
for fairly restricted situations.
The Penrose-like inequality for the TTS
is proved in Sect.~\ref{section4-4}.

\subsection{Setup and TTS condition}
\label{section4-1}

Consider an axisymmetric stationary spacetime $\mathcal{M}$.
There are two Killing vector fields:
One is the timelike Killing vector field $t^a$ and
the other is a spacelike Killing vector field $\phi^a$
that represents $U(1)$ isometry.
Since 
these two Killing fields have been proved
to commute \cite{Carter:1970},  
it is possible to adopt a $t=\mathrm{const.}$ slice on which $U(1)$ symmetry
becomes manifest. In addition, we require
the existence of two-dimensional surfaces that are
orthogonal to both $t^a$ and $\phi^a$
(called the $t$--$\phi$ orthogonality property).
The necessary and sufficient condition
for this requirement is proved to be \cite{Carter:1969}
\begin{equation}
  t^{a}T_{a[b} t_{c} \phi_{d]}\ =\ \phi^{a}T_{a{[}b} t_{c} \phi_{d]}\  =\ 0. 
\end{equation}
Physically, this condition means 
that matter, if it exists, is moving just in the $\phi^a$ direction. 
In this case, there exists the symmetry of the metric
under the transformation $t^a\to -t^a$ and $\phi^a\to -\phi^a$.

Since we consider the nonstatic case in this section,
the timelike Killing field is decomposed as
\begin{equation}
  t^a = \alpha n^a + \beta^a,
  \label{Stationary-Lapse-Shift}
\end{equation}
where $\alpha$ is the lapse function and $\beta^a$ is
the shift vector. Here, we consider the time slice
on which the shift vector $\beta^a$
is proportional to the Killing vector field $\phi^a$,
\begin{equation}
  \beta^a = -\omega\phi^a.
  \label{shift-phi-relation}
\end{equation}
Here, the quantity $\omega$ in Eq.~\eqref{shift-phi-relation}
corresponds to 
the angular velocity of the zero-angular-momentum
observers (ZAMOs).
Note that this condition is realized only on a special slice. 
For example, although the
Boyer--Lindquist coordinates of the Kerr spacetime
possess this property,  
other coordinates like the Kerr--Schild coordinates \cite{Kerr:1963}
or the Doran coordinates \cite{Doran:1999} do not satisfy this condition
because the shift vector has a radial component.

In this paper, we consider only axisymmetric TTSs
for a technical reason. If a TTS $S$ is axisymmetric,
the timelike unit normal $n^a$ to $\Sigma$ becomes
a tangent vector of $S$, because both $t^a$ and $\beta^a$
are tangent to $S$. Then, the outward unit normal $\hat{r}^a$
to $S$ becomes the outward unit normal to $S_0$ in $\Sigma$
as well. If $S$ is not axisymmetric, these properties
do not hold and the analysis becomes complicated.
The study of nonaxisymmetric TTSs in the stationary case
is left as a remaining problem.

Let us start the derivation of the TTS condition
in the axisymmetric stationary spacetime.
Because the induced metric of $S$ is given by
$P_{ab}=-n_an_b+h_{ab}$ in this setup, we have
\begin{equation}
  \bar{K}_{ab} = \frac{1}{2}\pounds_{\hat{r}}P_{ab} =
  -n_an_b\frac{\hat{r}^cD_c\alpha}{\alpha}
  +
  \frac{1}{2}\pounds_{\hat{r}}h_{ab}.
\end{equation}
On the other hand, rewriting the three-dimensional quantities
by the four-dimensional quantities, we have
\begin{equation}
k_{ab} = \frac12{}^{(3)}\pounds_{\hat{r}}h_{ab} =
  \frac12\pounds_{\hat{r}}h_{ab}
   +\frac12(n^d\nabla_d\hat{r}^c-\hat{r}^d\nabla_dn^c)(n_ah_{cb}+n_bh_{ca}).
\end{equation}
Putting these together, the decomposition
of the extrinsic curvature of $S$ is given as
\begin{equation}
    \bar{K}_{ab}\ =\ -n_an_b\frac{\hat{r}^cD_c\alpha}{\alpha}
    +k_{ab}
    -n_av_b-n_bv_{a}.
    \label{barKab-general}
\end{equation}
with
\begin{equation}
  v_a=\frac12 h_{ab}\pounds_n\hat{r}^b.
  \label{definition_va_1}
\end{equation}
Note that $v_a$ can be regarded as the cross-components of $\bar{K}_{ab}$, that is, 
$v_a={h_a}^b\bar{K}_{bc}n^c$.
Now, we recall the TTS condition \eqref{TTS-condition}.
As in the static case, the null tangent vector of $S$ is decomposed as
$k^a = n^a + s^a$ with a unit tangent vector $s^a$ of $S_0$.
Then, the TTS condition in this case is that
\begin{equation}
k_{ab}s^as^b +2v_bs^b\le\frac{\hat{r}^cD_c\alpha}{\alpha}
\label{TTS-condition-Stationary}
\end{equation}
holds for an arbitrary unit tangent vector $s^a$ of $S_0$.

This condition can be simplified further.
Because $S$ is stationary and axisymmetric,
the unit normal $\hat{r}^a$ of $S$
is Lie transported along the integral lines of the
Killing vectors:
$\pounds_{t}\hat{r}^a = {}^{(3)}\pounds_{\phi}\hat{r}^a = 0.$
Using this relation, the vector $v^b$ is rewritten as
\begin{equation}
  v_a = \frac{1}{2\alpha}h_{ab}{}^{(3)}\pounds_{\hat{r}}\beta^b
  =
  -\frac{1}{2\alpha}
  (\hat{r}^bD_b\omega){\phi}_a.
  \label{va_AxiStationary}
\end{equation}
On the other hand, the assumption of symmetry under the transformation
$t^a\to -t^a$ and $\phi^a\to -\phi^a$ (the $t$--$\phi$ orthogonality property)
implies that the geometry of the $t=\mathrm{const.}$ slice
has symmetry under $\phi^a\to -\phi^a$. Therefore,
the extrinsic curvature of $S_0$ satisfies
$k_{ab}\hat{\theta}^a{\phi}^b=0$, where $\hat{\theta}^a$
is the unit tangent vector of $S_0$ orthogonal to ${\phi}^a$.
Hence, introducing the tetrad basis as $(\mathbf{e}^1)_a=\phi_a/\phi$ and
$(\mathbf{e}^2)_a=\hat{\theta}_a$ (with $\phi=\sqrt{\phi^a\phi_a}$),  
$k_{ab}$ and $v_a$ are expressed as
\begin{subequations}
\begin{eqnarray}
  k_{ab} &=& k_1(\mathbf{e}^1)_a(\mathbf{e}^1)_b+
  k_2(\mathbf{e}^2)_a(\mathbf{e}^2)_b,
\label{kab-axisymmetric-stationary-tetrad}
  \\
  v_a &=& v_1(\mathbf{e}^1)_a,
\label{va-axisymmetric-stationary-tetrad}
\end{eqnarray}
\end{subequations}
with $v_1=-({\phi}/{2\alpha})(\hat{r}^bD_b\omega)$.
Setting $s_a=\cos\xi(\mathbf{e}^1)_a+\sin\xi(\mathbf{e}^2)_a$ and
substituting these expressions into \eqref{TTS-condition-Stationary},
the function
\begin{eqnarray}
  f(\cos\xi) = (k_2-k_1)\cos^2\xi-2v_1\cos\xi
  +\frac{\hat{r}^cD_c\alpha}{\alpha}-k_2
  \label{function-fx-axistationary-1}
\end{eqnarray}
must be nonnegative for arbitrary $0\le \xi< 2\pi$.
This condition can be re-expressed in terms of the
relation between the coefficients
(see Appendix \ref{appendix-b} for a derivation),
and we have the following result:
%
\begin{proposition}
  The necessary and sufficient condition
  for $S$ to be an axisymmetric TTS in an axisymmetric
  stationary spacetime with the $t$--$\phi$ orthogonality property
  is that one of the following
three conditions is satisfied at any point on $S_0$:
\begin{subequations}
\begin{eqnarray}
    & \textrm{(i)} &
  \frac{\hat{r}^cD_c\alpha}{\alpha}-k_1\ge |2v_1|> 2(k_2-k_1)>0;
  \label{TTS-condition-AxiStationary-1}\\
    & \textrm{(ii)} &
    \frac{\hat{r}^cD_c\alpha}{\alpha}-k_2\ge \frac{v_1^2}{k_2-k_1} 
  \qquad\textrm{and}\qquad
  k_2-k_1\ge |v_1|;
  \label{TTS-condition-AxiStationary-2}\\
  & \textrm{(iii)} &
  \frac{\hat{r}^cD_c\alpha}{\alpha}- |2v_1|\ge k_1\ge k_2.
  \label{TTS-condition-AxiStationary-3}
\end{eqnarray}
\end{subequations}
\label{proposition-3}
\end{proposition}
%

Similarly to the static case, this condition
can be derived by directly studying the geodesic equations.
This is demonstrated in Appendix~\ref{appendix-a2}.
For a convex TTS, the conditions $k_1\ge 0$ and $k_2\ge 0$
are additionally required.

\subsection{Examples in a Kerr spacetime}
\label{section4-2}

Here, we explicitly demonstrate that TTSs 
exist in a stationary spacetime using the Kerr spacetime
as an example. We start with a sufficiently general metric
of axisymmetric stationary spacetimes
possessing the $t$--$\phi$ orthogonality property, 
\begin{equation}
  ds^2=-{\alpha}^2dt^2+\gamma^2(d\phi-\omega dt)^2 +\varphi^2dr^2+\psi^2d\theta^2,
  \label{Axisymmetric-Stationary-Metric}
\end{equation}
where all metric functions depend on $r$ and $\theta$ only.
Here, $\alpha$ and $\omega$ are same as those
given in Eqs.~\eqref{Stationary-Lapse-Shift} and \eqref{shift-phi-relation}
(i.e., the lapse function and the ZAMO angular velocity).
Here, we consider the condition that the surface $r=\mathrm{const.}$ becomes
a TTS. The tetrad components of the extrinsic
curvature $k_{ab}$ and the vector $v_a$
defined in Eqs.~\eqref{kab-axisymmetric-stationary-tetrad}
and \eqref{va-axisymmetric-stationary-tetrad}
are calculated as
\begin{equation}
  k_1 = \frac{\gamma_{,r}}{\varphi\gamma}, \qquad
  k_2 = \frac{\psi_{,r}}{\varphi\psi},\qquad
  v_1 = -\frac{\gamma\omega_{,r}}{2\varphi\alpha},
  \label{MetricFormula-k1-k2-v1}
\end{equation}
and the relation $\hat{r}^cD_c\alpha/\alpha = \alpha_{,r}/(\varphi\alpha)$
is also used.

The metric functions of a Kerr spacetime
in the Boyer--Lindquist coordinates (with $G=1$) are given by
\begin{equation}
    {\alpha}^2 = \frac{\Delta\Sigma}{\mathcal{A}},\qquad
    \gamma^2 = \frac{\mathcal{A}\sin^2\theta}{\Sigma},\qquad
    \omega = \frac{2Mar}{\mathcal{A}},\qquad
    \varphi^2 =\frac{\Sigma}{\Delta},\qquad
    \psi^2 = \Sigma,
\end{equation}
with
\begin{subequations}
  \begin{equation}
    \Sigma = r^2+a^2\cos^2\theta,
    \qquad
    \Delta = r^2+a^2-2Mr,
    \end{equation}
\begin{equation}
  \mathcal{A} = (r^2+a^2)^2-\Delta a^2\sin^2\theta.
\end{equation}
\end{subequations}
Here, $M$ is the ADM mass and $a$ is the rotation parameter
that is related to the ADM angular momentum as $J=Ma$.
The event horizon is located at $r=r_+=M+\sqrt{M^2-a^2}$.
We consider the parameter region $0<a\le M$, i.e., a nonstatic spacetime
with an event horizon.

%
\begin{figure}[tb]
\centering
\includegraphics[width=0.5\textwidth,bb=0 0 259 172]{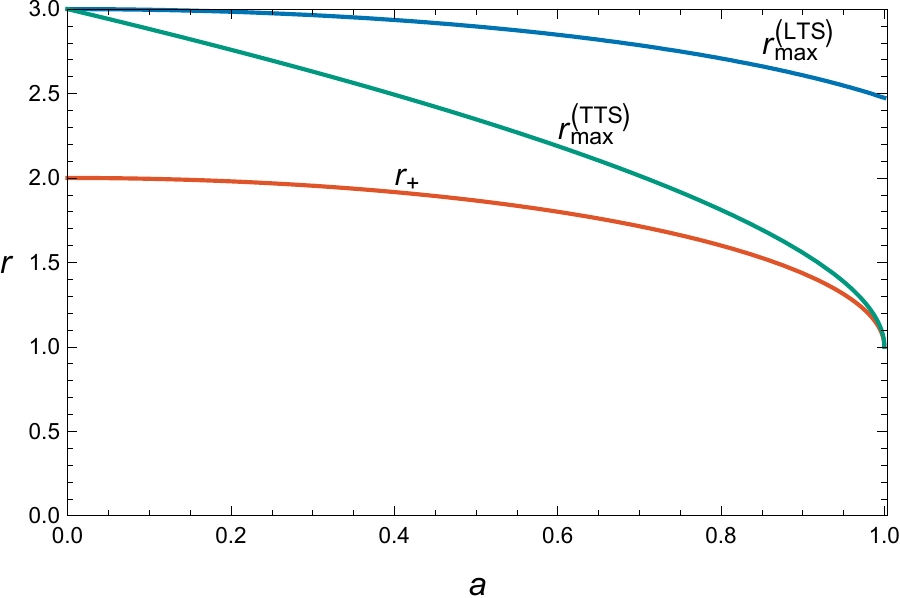}
\caption{ The behavior of $r^{\rm (TTS)}_{\rm max}$ and
  $r^{\rm (LTS)}_{\rm max}$ as functions of $a$. 
  The unit of the length is $M$.
}
\label{Figure-rmaxTTS-rmaxLTS}
\end{figure}
%

For a Kerr spacetime, it is easy to check 
$k_2>k_1$, which is natural because
the Kerr black hole has oblate geometry.
Although the calculation is more tedious,
one can show that $v_1^2>(k_2-k_1)^2$ holds.
Therefore, we consider case (i) of Proposition \ref{proposition-3},
and the inequality ${\hat{r}^cD_c\alpha}/{\alpha}-k_1\ge |2v_1|$
in \eqref{TTS-condition-AxiStationary-1} determines the
range of $r$. The squared inequality
gives the condition
\begin{equation}
r(r-3M)^2\ge 4a^2M
\end{equation}
on the equatorial plane $\theta=\pi/2$ (the condition is weaker 
for other $\theta$ values). There are two domains satisfying
this inequality, and just the inner one satisfies
the nonsquared inequality 
(in the outer domain, ${\hat{r}^cD_c\alpha}/{\alpha}-k_1$ becomes negative).
In this way, we find that the surface $r=\textrm{const.}$
is a TTS for
\begin{equation}
  r_+<r<r^{\rm (TTS)}_{\rm max} = 2M\left\{
1+\cos\left[\frac{2}{3}\arccos\left(-\frac{a}{M}\right)\right]
  \right\}.
\end{equation}
The behavior of $r^{\rm (TTS)}_{\rm max}$ is shown as
a function of $a/M$ in Fig.~\ref{Figure-rmaxTTS-rmaxLTS}.
Note that the value of $r^{\rm (TTS)}_{\rm max}$
corresponds to the radius of the circular orbit of a photon
closest to the black hole (e.g., p.~73 of \cite{Frolov:1998}).
Therefore, our result is reasonable
because only photons
propagating on the equatorial plane in the direction of
the black hole rotation propagate on the surface $r=r^{\rm (TTS)}_{\rm max}$,
and all other photons initially moving in the tangent direction to the
surface will fall into the black hole.

Let us also look at LTSs in the Kerr spacetime.
A surface $r=\mathrm{const.}$ becomes an LTS
when $k_{,r}\ge 0$  is satisfied at every point.
This condition becomes
strictest on the equatorial plane $\theta=\pi/2$,
and there exists a maximum radius $r^{\rm (LTS)}_{\rm max}$ such that
an $r=\mathrm{const}.$ surface becomes
an LTS for $r_+\le r\le r^{\rm (LTS)}_{\rm max}$.
Unfortunately, no simple analytic formula
for $r^{\rm (LTS)}_{\rm max}$ seems to exist, 
unlike the TTS case. The behavior
of $r^{\rm (LTS)}_{\rm max}$ is shown in Fig.~\ref{Figure-rmaxTTS-rmaxLTS}.
Although $r^{\rm (LTS)}_{\rm max}=r^{\rm (TTS)}_{\rm max}=3M$ at $a/M=0$,  
LTSs distribute in a broader region compared to TTSs
when $a/M$ is large. 
While $r=r^{\rm (TTS)}_{\rm max}$ indicates the inner edge of
the photon region, $r=r^{\rm (LTS)}_{\rm max}$ is located
in the middle of the photon region.
In this sense, the TTS and the LTS are different indicators
for strong gravity regions,
and the TTS is a stricter one compared to the LTS.

\subsection{Connection to the LTSs}
\label{section4-3}

Here, we try to derive the condition that a convex TTS
becomes an LTS using the Einstein equation.
Since $\pounds_th_{ab}=0$ in a stationary spacetime, we have
\begin{equation}
  K_{ab} =-\frac{1}{2\alpha}\left(D_a\beta_b+D_b\beta_a\right).
  \label{Evolution1-Stationary}
\end{equation}
Using the projection \eqref{Energy-Momentum-Tensor-Projection}
of the energy momentum tensor,
the Einstein equations in the $3+1$ split form are
\begin{subequations}
  \begin{eqnarray}
    {}^{(3)}R+K^2-K_{ab}K^{ab} &=& 16\pi G\rho,
    \label{Hamiltonian-Stationary}\\
    D_a{K^a}_b -D_bK &=&-8\pi GJ_b,
    \label{Momentum-Stationary}
    \\
    {}^{(3)}R_{ab}+K_{ab}K-2K_{ac}{K^{c}}_b
    &=&\frac{1}{\alpha}{}^{(3)}\pounds_\beta K_{ab}
    +\frac{1}{\alpha}D_aD_b\alpha
    \nonumber\\
    && \qquad
    + 8\pi G\left[S_{ab} + \frac12 q_{ab}(\rho-{S^c}_c)\right].
    \label{Evolution2-Stationary}
\end{eqnarray}
\end{subequations}
Here, Eqs.~\eqref{Hamiltonian-Stationary} and \eqref{Momentum-Stationary}
are the Hamiltonian and momentum constraints, respectively,
and Eq.~\eqref{Evolution2-Stationary} is the evolution equation
with $\pounds_tK_{ab}=0$.
Because of the axisymmetry of the space $\Sigma$,
the Lie derivatives with respect to $\phi^a$ of the geometric quantities
become zero:
\begin{equation}
  {}^{(3)}\pounds_\phi\alpha\ =\ {}^{(3)}\pounds_{\phi}\beta^a\ =\
  {}^{(3)}\pounds_{\phi}q_{ab}\ =\ {}^{(3)}\pounds_{\phi}K_{ab}\ =\ 0.
\end{equation}
Here, ${}^{(3)}\pounds_{\phi}q_{ab}=0$ is equivalent to the 
Killing equation of $\phi^a$ on $\Sigma$, $D_a\phi_b+D_b\phi_a=0$.
From Eq.~\eqref{shift-phi-relation},
${}^{(3)}\pounds_{\phi}\beta^a=0$ is equivalent to $\phi^bD_b\omega=0$,
and 
\begin{equation}
  K_{ab} = \frac{1}{2\alpha}
  \left(\phi_aD_b\omega+\phi_bD_a\omega\right). \label{KabpDo}
\end{equation}
Taking the trace of this relation, we have $K=0$, i.e. 
the spacelike hypersurface $\Sigma$ is maximally sliced.
Substituting these relations,
the Hamiltonian constraint~\eqref{Hamiltonian-Stationary}
and the evolution equation~\eqref{Evolution2-Stationary}
become
  \begin{subequations}
  \begin{eqnarray}
{}^{(3)}R & = &
    \frac{\phi^2}{2\alpha^2}(D_b\omega D^b\omega) + 16\pi G\rho,
    \label{Hamiltonian-AxiStationary}\\
{}^{(3)}R_{ab} & = &
          \frac{1}{2\alpha^2}
          \left[\phi_a\phi_b(D_c\omega D^c\omega)-\phi^2D_a\omega D_b\omega\right]
          \nonumber\\
          &&\qquad\qquad\qquad
          +\frac{1}{\alpha}D_{a}D_{b}\alpha
          +8\pi G\left[S_{ab} + \frac12q_{ab}(\rho-{S^c}_c)\right],
          \label{Evolution2-AxiStationary}
\end{eqnarray}
\end{subequations}
with $\phi^2=\phi^a\phi_a$. The trace of
Eq.~\eqref{Evolution2-AxiStationary} gives
\begin{equation}
    \frac{1}{\alpha}D^2\alpha = \frac{\phi^2}{2\alpha^2}(D_c\omega D^c\omega)
    +4\pi G(\rho+{S^c}_c).
    \label{Einstein-Axistationary-3}
\end{equation}

Consider a convex TTS whose spatial section is $S_0$.
We rewrite the formula~\eqref{LTS-Ricci-equation}
for $\hat{r}^aD_ak$ of $S_0$
using the Gauss equation \eqref{Gauss-SpacelikeHypersurface}.
The quantity ${}^{(3)}R_{ab}\hat{r}^a\hat{r}^b$ in
Eq.~\eqref{Gauss-SpacelikeHypersurface}
is rewritten by
\begin{equation}
{}^{(3)}R_{ab}\hat{r}^a\hat{r}^b=    8\pi G(\rho+P_r)  
    -\frac{1}{\alpha}\mathcal{D}^2\alpha
    -k\frac{\hat{r}^cD_c\alpha}{\alpha}
    +\frac{\phi^2}{2\alpha^2}(\mathcal{D}_c\omega \mathcal{D}^c\omega),
    \label{rho-plus-Pr-AxiStationary}
\end{equation} 
which is obtained by multiplying 
Eq.~\eqref{Evolution2-AxiStationary} by $\hat{r}^a\hat{r}^b$ and rewriting with
Eq.~\eqref{Einstein-Axistationary-3}.
Here, $P_r$ is the pressure in the radial direction introduced in Eq.~\eqref{Radial-Pressure}.
The result is (see also Appendix C)
\begin{equation}
  \hat{r}^aD_ak
 \ =\
-8\pi G(\rho+P_r) 
    +\frac{1}{\alpha}\mathcal{D}^2\alpha
    +k\frac{\hat{r}^cD_c\alpha}{\alpha}
    -k_{ab}k^{ab}
    -\frac{\phi^2}{2\alpha^2}(\mathcal{D}_c\omega \mathcal{D}^c\omega)
    -\frac{1}{\varphi}\mathcal{D}^2\varphi. \label{rDtrk2}
\end{equation}
We would like to find the condition that the positivity of $\hat{r}^aD_ak$
is guaranteed. Similarly to the static case, we impose $\varphi=\alpha$
and $\rho+P_r=0$. Then, if the condition
\begin{equation}
k\frac{\hat{r}^cD_c\alpha}{\alpha}
-k_{ab}k^{ab}\ge \frac{\phi^2}{2\alpha^2}(\mathcal{D}_c\omega \mathcal{D}^c\omega)
\label{sufficient-condition-Axistationary-LTS}
\end{equation}
is satisfied, a TTS is guaranteed to be an LTS. However, unlike
in the static case, the existence of the right-hand side
makes it difficult to prove this inequality.

%
\begin{figure}[tb]
\centering
\includegraphics[width=0.4\textwidth,bb=0 0 260 243]{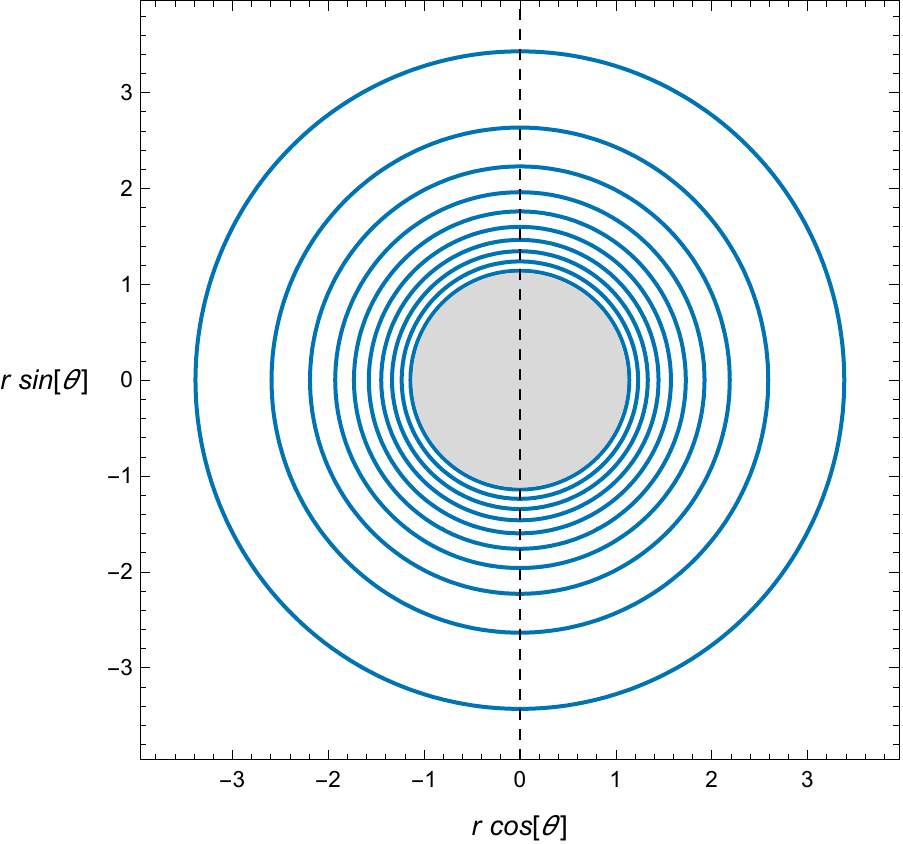}
\caption{
  Contour surfaces of the ZAMO angular velocity $\omega$
  for $\omega/\Omega_{\rm H}=n/10$ with $n=1, . . . ,10$, where $\Omega_{\rm H}$
  is the angular velocity of the event horizon,
  in the $(r\cos\theta,r\sin\theta)$-plane of a Kerr spacetime
  with $a/M=0.99$. The unit of the length is $M$.
}
\label{ZAMO-equiAV}
\end{figure}
%

Instead of making a general argument, we consider
surfaces each of which is given by $\omega=\mathrm{const.}$
The $\omega=\mathrm{const.}$ surface consists of trajectories
of ZAMO observers with the same angular velocity $\omega$.
Examples of contour surfaces of the ZAMO angular velocity
are depicted in Fig.~\ref{ZAMO-equiAV} in the case of a Kerr spacetime
with $a/M=0.99$. These surfaces are approximately the same as 
the $r=\mathrm{const.}$ surfaces.
For this surface, the right-hand side of the inequality
\eqref{sufficient-condition-Axistationary-LTS}
vanishes. By setting $s^a=\pm (\mathbf{e}^1)^a$ and $(\mathbf{e}^2)^a$
in the TTS condition \eqref{TTS-condition-Stationary}, we have
${\hat{r}^cD_c\alpha}/{\alpha}\ge k_1+2|v_1|\ge k_1$ and
${\hat{r}^cD_c\alpha}/{\alpha}\ge k_2$. 
These two inequalities are expressed as
${\hat{r}^cD_c\alpha}/{\alpha}\ge k_{\rm L}$
by setting $k_{\rm L}:=\mathrm{max}(k_1,k_2)$
and $k_{\rm S}:=\mathrm{min}(k_1,k_2)$.
Using this, the relation
\begin{equation}
k\frac{\hat{r}^cD_c\alpha}{\alpha}
-k_{ab}k^{ab} =
k_{\rm S}(k_{\rm L}-k_{\rm S})\ge 0
\end{equation}
holds for a convex TTS. To summarize, we have proved the following:
%
\begin{proposition}
  If a contour surface of the ZAMO angular velocity
  is a convex TTS and $\rho+P_r=0$ on it, it is an LTS as well.
  \label{proposition-4}
\end{proposition}
%

Compared to the static case, the argument here is fairly restricted,
not because a convex TTS is not an LTS
in many situations, but because the method here does not work sufficiently.
As we have seen in the Kerr case, the TTSs given by $r=\mathrm{const}.$
surfaces are simultaneously
LTSs for all $0\le a/M\le 1$. We expect that a better method may
establish the connection between the TTS and the LTS
in the axisymmetric stationary case more firmly.
This is left as a remaining problem.

\subsection{Area bound for TTSs}
\label{section4-4}

If we are only interested in the area bound by the Penrose-like
inequality, it is possible to relax the condition
of Proposition \ref{proposition-4} in a similar manner to the static case
in Sect.~\ref{section3-3}. Namely, we consider the condition
that leads to the inequality \eqref{integral-k2dA}
on a convex TTS $S$ without using the concept of the LTS.
Eliminating ${}^{(3)}R$ and ${}^{(3)}R_{ab}\hat{r}^a\hat{r}^b$
from Eqs.~\eqref{Gauss-SpacelikeHypersurface},
\eqref{Hamiltonian-AxiStationary}, and \eqref{rho-plus-Pr-AxiStationary},
we have (see also Appendix C)
\begin{equation}
  {}^{(2)}R = -16\pi GP_r
  +\frac{2}{\alpha}{\mathcal{D}^2\alpha}
  +2k\frac{\hat{r}^aD_a\alpha}{\alpha}
  +k^2-k_{ab}k^{ab}
  +\frac{\phi^2}{2\alpha^2}\left[
(\hat{r}^aD_a\omega)^2-(\mathcal{D}_a\omega\mathcal{D}^a\omega)
    \right].
  \label{2D-RicciScalar-AxiStationary}
\end{equation}
The inequality ${\hat{r}^cD_c\alpha}/{\alpha}\ge k_{\rm L}$ implies
\begin{eqnarray}
2k\frac{\hat{r}^aD_a\alpha}{\alpha}
+k^2-k_{ab}k^{ab}
\ge 
\frac{3}{2}k^2 + \frac12(k_{\rm L}+3k_{\rm S})(k_{\rm L}-k_{\rm S})
\ge 
\frac{3}{2}k^2 
\end{eqnarray}
for a convex TTS (a nonconvex TTS also satisfies this inequality
if $k_{\rm S}$ is within the range $k_{\rm S}\ge -k_{\rm L}/3$).
Assuming $P_r\le 0$ and integrating the relation \eqref{2D-RicciScalar-AxiStationary},
we have
\begin{equation}
  \int_{S_0}{}^{(2)}RdA
  \ge
  \int_{S_0}
  \left\{  \frac{3}{2}k^2
  +\frac{2\mathcal{D}_a\alpha\mathcal{D}^a\alpha}{\alpha^2}
  +\frac{\phi^2}{2\alpha^2}\left[
(\hat{r}^aD_a\omega)^2-(\mathcal{D}_a\omega\mathcal{D}^a\omega)
    \right]
  \right\}dA.
\end{equation}
If the right-hand side is positive, $S_0$ has topology $S^2$ and  
the left-hand side becomes $\int_{S_0}{}^{(2)}RdA=8\pi$
because of the Gauss--Bonnet theorem.
Therefore, we have the following result:
%
\begin{theorem}
  The time cross section of an axisymmetric convex TTS,
  $S_0$, has topology $S^2$  and
  satisfies the Penrose-like inequality
  $A_0\le 4\pi (3GM)^2$ if $P_r\le 0$ and
\begin{equation}
  \frac{4}{\phi^{2}}(\mathcal{D}_a\alpha\mathcal{D}^a\alpha)
  + (\hat{r}^aD_a\omega)^2
  \ge
  (\mathcal{D}_a\omega\mathcal{D}^a\omega)
  \label{AxiStationary-Penrose-Assumption}
\end{equation}
  holds on $S_0$, $k>0$ at least at one point on $S_0$, and ${}^{(3)}R$
  is nonnegative in the outside region in an
  asymptotically flat axisymmetric stationary spacetime
  with the $t$--$\phi$ orthogonality property.
\label{theorem-3}
\end{theorem}
%

Here, we discuss to what extent the condition~\eqref{AxiStationary-Penrose-Assumption}
is strong.
Since $\phi^a\mathcal{D}_a\omega=0$, the right-hand side is
rewritten as $(\hat{\theta}^aD_a\omega)^2$. This quantity
is zero at the symmetry axis, and also on an equatorial plane
(if it exists). Therefore, the condition \eqref{AxiStationary-Penrose-Assumption}
is satisfied at least at these two locations.
Furthermore, since the ZAMO angular momentum $\omega$
coincides with the (constant) horizon angular velocity
$\Omega_{\rm H}$ on the event horizon, 
the condition \eqref{AxiStationary-Penrose-Assumption}
must be satisfied at least on surfaces sufficiently
close to the event horizon. For this reason, the condition \eqref{AxiStationary-Penrose-Assumption}
does not restrict the situation strongly, and
hence many TTSs satisfying this condition should exist.
In fact, we can check that this condition is satisfied
by arbitrary $r=\mathrm{const.}$ surfaces of a Kerr spacetime.
Therefore, Theorem \ref{theorem-3} is expected to have sufficient generality.

%
%
\section{Summary and discussion}
\label{section5}

In this paper, we have defined a new concept,
the transversely trapping surface, as a generalization
of static photon surfaces.
Its definition was introduced in Sect.~\ref{section2-1}
(Definition~\ref{definition-1}), and 
the condition for a surface $S$
to be a TTS is mathematically expressed as Eq.~\eqref{TTS-condition}
in Proposition~\ref{proposition-1}.
The properties of the TTS in static spacetimes
were studied in Sect.~\ref{section3}.
There, a TTS is proved to be a loosely trapped surface as 
defined in our previous paper \cite{Shiromizu:2017}
(see Definition~\ref{definition-2} of this paper)
at the same time under certain conditions (Proposition \ref{proposition-2} in Sect.~\ref{section3-2}).
The area of a TTS is shown to
satisfy the Penrose-like inequality~\eqref{Penroselike-inequality}
under some generic conditions (Theorem \ref{theorem-2} in Sect.~\ref{section3-3}).

In Sect.~\ref{section4}, we studied TTSs
in axisymmetric stationary spacetimes.
Because of a technical reason, we considered
axisymmetric TTSs in spacetimes with the $t$--$\phi$ orthogonality
property. The TTS condition in this setup was summarized
in a concise form (Proposition \ref{proposition-3} in Sect.~\ref{section4-1}).
It was explicitly shown that TTSs exist in a Kerr spacetime
(Sect.~\ref{section4-2}). As for the connection to the LTS,  
we have shown that TTSs given by contour surfaces of the ZAMO angular velocity
are LTSs at the same time under certain conditions
(Proposition \ref{proposition-4} in Sect.~\ref{section4-3}).
This fairly restricted argument is due to a technical reason,
and generalization is left as an open problem. 
However, we have proved
the Penrose-like inequality~\eqref{Penroselike-inequality}
under fairly general situations
(Theorem \ref{theorem-3} in Sect.~\ref{section4-4}).

We have established that 
the area of a TTS is bounded from above by the area of a photon sphere
with the same mass in quite generic static/stationary situations.
This is a natural result, because 
if photons propagating in the transverse direction to the source
are trapped, such a region must be compact
so that gravity is sufficiently strong.

It is interesting to list spacetimes
possessing TTSs. Black hole spacetimes generally
possess TTSs around their horizons. In addition to
the vacuum black holes, 
black holes surrounded by ring-shaped matter \cite{Ansorg:2005}
and those with scalar or proca
hairs \cite{Herdeiro:2014,Herdeiro:2015,Herdeiro:2016} also possess TTSs.
There are a few examples of spacetimes without horizons
where TTSs are present.
One is a spherically symmetric
star composed of incompressible fluid (e.g., \cite{Wald}):
the radius of such a star $R$ can be as small as $8/3\le R/M\le 3$ without
violating the dominant energy condition.
Another example is the soliton-like structure of a complex massive
scalar field, the so-called boson star.
It has been reported in \cite{Horvat:2013} that
a boson star can have a photon sphere (and, therefore, TTSs), 
although the binding energy is not negative for such situations 
if the scalar field is minimally coupled to gravity.
Other rather exotic examples are spacetimes with naked singularities,
as studied in \cite{Virbhadra:2002,Bozza:2002,Virbhadra:2007,Sahu:2012,Sahu:2013}.

In this paper, we restricted the definition
of the TTS as a static/stationary surface
in a static/stationary spacetime.
It would be possible to generalize this definition
to more general surfaces in dynamical spactimes.
For example, one may define the generalized TTS
as an arbitrary spatially bounded timelike surface $S$
on which the TTS condition $\bar{K}_{ab}k^ak^b\le 0$
holds for arbitrary null tangent vectors $k^a$ everywhere.
Taking account of this possibility, in Appendix~\ref{appendix-c}
we present some useful geometric formulas
in the setup given by Fig.~\ref{schematic-TTS} but without
assuming the timelike Killing symmetry.
As special cases, the formulas in the main text are 
rederived as a consistency check.

There are many possible directions of extensions of our current work.
Some examples are listed here. 
First, for axisymmetric spacetimes with nonvanishing global angular momentum $J$,
a generalization of the Penrose inequality ,
\begin{equation}
A_{\rm H}\le 8\pi G^2\left(M^2+\sqrt{M^4-\frac{J^2}{G^2}}\right),
\end{equation}
has been conjectured for apparent horizons \cite{Dain:2002}.
Namely, the bound of the apparent horizon area
is expected to become stronger as $J$
is increased.
From the study of a Kerr spacetime in Sect.~\ref{section4-2} of this paper,
the same tendency is also expected for TTSs.
Looking for a stronger bound of the TTS area
depending on $J$ is an interesting problem.
Furthermore, an inequality of a different type,
\begin{equation}
8\pi G|J_q|\le A_{\rm H},
\end{equation}
has been proved for an apparent horizon in an axisymmetric
spacetime under some generic
conditions, where $J_q$ is an appropriately defined
quasilocal angular momentum \cite{Dain:2011}.
The lower bound of the TTS area in terms of $J_q$
is also an interesting issue.

Second, there is another formulation for compactness
of an apparent horizon, the hoop conjecture \cite{Thorne:1972},
which states that
black holes with horizons form when and only when a mass
$M$ gets compacted into a region whose circumference in
every direction is bounded by $C\lesssim 2\pi (2GM)$.
The important claim of this conjecture is that an apparent horizon
does not become arbitrarily long in one direction, 
and this property was shown to hold  
in several examples (e.g., \cite{Yoshino:2007}).
It would be interesting to test whether TTSs also satisfy
this property. We expect that an argument
such that the circumference of a TTS is bounded
as $C\lesssim 2\pi (3GM)$ could be made.

Finally, since a TTS suggests that gravity
is strong there, ordinary matter would not be able to support
itself inside of the TTS. Therefore, it would be possible
to show the existence of a horizon inside of a TTS
under some conditions.
In the spherically symmetric case,
it has been shown that a perfect fluid star
consisting of polytropic balls cannot possess a photon sphere \cite{Saida:2015}. 
We expect that it would be possible to
generalize such an argument for nonspherical TTSs 
applying the methods for proving singularity theorems.

\ack

The work of H. Y. was in part supported by a Grant-in-Aid for
Scientific Research (A) (No. 26247042)
from the Japan Society for the Promotion of Science (JSPS).
K. I. is supported by a JSPS Grant-in-Aid for Young Scientists (B) (No. 17K14281).
T. S. is supported by a Grant-in-Aid for Scientific Research (C) (No. 16K05344)
from JSPS.

\appendix

%
%
\section{TTS conditions from geodesic equations}

The purpose of this appendix is to demonstrate that
the TTS conditions in the static case and in the
axisymmetric stationary case can be derived by directly examining
the geodesic equations. 

\subsection{Static spacetimes}
\label{appendix-a1}

The metric of a static spacetime with static slices $t=\mathrm{const.}$ 
is given by
\begin{equation}
  ds^2 = -\alpha^2dt^2+\varphi^2dr^2 + h_{IJ}dx^Idx^J.
\end{equation}
Consider a null geodesic
whose tangent vector is $k^a=(\dot{t}, \dot{r}, \dot{x}^I)$ with $I=1$
and $2$, where the dot denotes the derivative with respect to the
affine parameter.
For a surface $S$ given by $r=\mathrm{const.}$ to be a TTS,
$\ddot{r}\le 0$ must hold 
for arbitrary $k^a$ with $\dot{r}=0$
at arbitrary points on $S$.
The radial equation is given by
\begin{equation}
  \ddot{r}=-\frac{\alpha\alpha_{,r}}{\varphi^2}\dot{t}^2-\frac{\varphi_{,r}}{\varphi}\dot{r}^2
  -\frac{2\varphi_{,I}}{\varphi}\dot{x}^I\dot{r} +\frac{h_{IJ,r}}{2\varphi^2}\dot{x}^I\dot{x}^J.
\end{equation}
Imposing $\dot{r}=0$ and eliminating $\dot{t}$ using
the null condition $k_ak^a=0$, we have
\begin{equation}
  \ddot{r}=\frac{1}{\varphi}
  \left(k_{IJ}-\frac{\hat{r}^aD_a\alpha}{\alpha}h_{IJ}\right)\dot{x}^I\dot{x}^J,
\end{equation}
where we used $\hat{r}^aD_a\alpha=\alpha_{,r}/\varphi$
and $k_{IJ}=h_{IJ,r}/{2\varphi}$. In order that $\ddot{r}\le 0$
holds for arbitrary $\dot{x}^I$, the matrix $\sigma_{ab}$
defined in \eqref{matrix-sigma} must have two nonpositive
eigenvalues. Therefore, we obtain the same TTS condition.

\subsection{Axisymmetric stationary spacetimes}
\label{appendix-a2}

We study the geodesic equation of an axisymmetric stationary spacetime
with the $t$--$\phi$ orthogonality property 
using the general metric~\eqref{Axisymmetric-Stationary-Metric}.
Similarly to the static case, we consider a null geodesic
with a tangent vector $k^a=(\dot{t}, \dot{r}, \dot{\theta}, \dot{\phi})$,
and require a surface $S$ given by $r=\mathrm{const.}$ to be a TTS.
Then, $\ddot{r}\le 0$ must hold 
for arbitrary $k^a$ with $\dot{r}=0$
at arbitrary points on $S$.
The radial equation is given by
\begin{equation}
  -\varphi^2\ddot{r}
  =
  {\alpha}{\alpha}_{,r}\dot{t}^2
  +\varphi\varphi_{,r}\dot{r}^2
  +2\varphi\varphi_{,\theta}\dot{r}\dot{\theta}
  -\psi\psi_{,r}\dot{\theta}^2
  -\gamma\gamma_{,r}(\dot{\phi}-\omega\dot{t})^2
  +\gamma^2\omega_{,r}(\dot{\phi}-\omega\dot{t})\dot{t}.
\end{equation}
Here, we impose $\dot{r}=0$ and rewrite with the null condition $k_ak^a=0$.
The difference from the static case is that eliminating
$\dot{t}$ causes the appearance of square roots in the equation
and makes the analysis complicated. Instead,
we eliminate $\dot{\theta}$ as
\begin{equation}
  -\varphi^2\ddot{r} =
  \left(\frac{{\alpha}_{,r}}{{\alpha}}-\frac{\psi_{,r}}{\psi}\right)
  ({\alpha}\dot{t})^2
  +\frac{\gamma\omega_{,r}}{{\alpha}}({\alpha}\dot{t})[\gamma(\dot{\phi}-\omega\dot{t})]
  +\left(\frac{\psi_{,r}}{\psi}-\frac{\gamma_{,r}}{\gamma}\right)
  [\gamma(\dot{\phi}-\omega\dot{t})]^2.
 \label{radial-geodesic-axisymmetric-stationary}
\end{equation}
Here, the left-hand side must be nonnegative for arbitrary $\alpha\dot{t}$
and $\gamma(\dot{\phi}-\omega\dot{t})$ satisfying the constraint 
$(\alpha\dot{t})^2\ge [\gamma(\dot{\phi}-\omega\dot{t})]^2$.
Introducing
\begin{equation}
x=\frac{\gamma(\dot{\phi}-\omega\dot{t})}{\alpha\dot{t}},
\end{equation}
Eq.~\eqref{radial-geodesic-axisymmetric-stationary}
is rewritten as $-\varphi\ddot{r} = (\alpha\dot{t})^2f(x)$ with
\begin{equation}
f(x) = \frac{1}{\varphi}\left(\frac{{\alpha}_{,r}}{{\alpha}}-\frac{\psi_{,r}}{\psi}\right)
  +\frac{\gamma\omega_{,r}}{{\varphi\alpha}}x
  +\frac{1}{\varphi}\left(\frac{\psi_{,r}}{\psi}-\frac{\gamma_{,r}}{\gamma}\right)x^2.
  \label{function-fx-geodesic-axistationary}
\end{equation}
Therefore, this function $f(x)$ must be nonnegative in the interval
$[-1,1]$.
Using the relation \eqref{MetricFormula-k1-k2-v1} between the tetrad components
of $k_{ab}$ and $v_a$ and the metric functions,
we find that the function $f(x)$ here
is exactly equivalent to Eq.~\eqref{function-fx-axistationary-1}
or Eq.~\eqref{function-fx-axistationary-2}.

%
%
\section{Derivation of the conditions
  \eqref{TTS-condition-AxiStationary-1}--\eqref{TTS-condition-AxiStationary-3}}
\label{appendix-b}

%
\begin{figure}[tb]
\centering
\includegraphics[width=0.8\textwidth,bb=0 0 579 166]{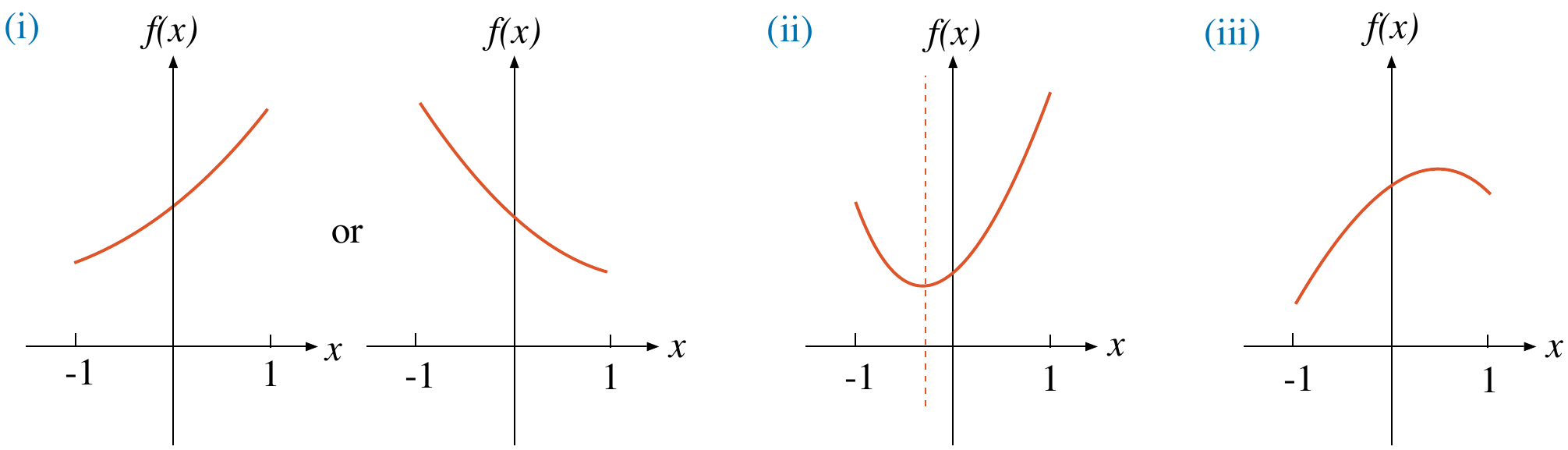}
\caption{
  The three cases (i)--(iii) that $f(x)$ becomes nonpositive
  in the region $-1\le x\le 1$.
}
\label{parabolic}
\end{figure}
%

Here, we present the detailed derivation of
the TTS conditions \eqref{TTS-condition-AxiStationary-1}--\eqref{TTS-condition-AxiStationary-3}
for axisymmetric stationary spacetimes.
The problem is reduced to clarifying the condition that
\begin{equation}
  f(x) = (k_2-k_1)x^2-2v_1x
  +\frac{\hat{r}^cD_c\alpha}{\alpha}-k_2
  \label{function-fx-axistationary-2}
\end{equation}
becomes nonnegative in the interval $[-1,1]$.
This parabola has the axis at $x=x_{\rm a}$ with
\begin{equation}
x_{\rm a} = \frac{v_1}{k_2-k_1},
\end{equation}
and there, it takes the value 
\begin{equation}
  f(x_{\rm a}) = \frac{\hat{r}^cD_c\alpha}{\alpha}-k_2 - \frac{v_1^2}{k_2-k_1}.
\end{equation}
At the endpoints of the interval $[-1,1]$, $f(x)$ satisfies
\begin{equation}
  f(\pm 1) 
  \ge \frac{\hat{r}^cD_c\alpha}{\alpha}-k_1- 2|v_1|.
\end{equation}
The condition can be studied by dividing it into
the three cases as indicated in Fig.~\ref{parabolic}:
\begin{itemize}
\item[(i)] The case that
$k_2> k_1$ and there is no
axis in the interval $[-1,1]$, i.e., $|x_{\rm a}|> 1$.
Since $f(x)$ takes the minimum value at
one of the endpoints, we require
$f(\pm 1)\ge 0$.
This leads to the condition~\eqref{TTS-condition-AxiStationary-1}.

\item[(ii)] The case that
$k_2> k_1$ and the axis exists
in the interval $[-1,1]$, i.e., $|x_{\rm a}|\le 1$.
Since $f(x)$ takes the minimum value at
the axis, we require
$f(x_{\rm a})\ge 0$.
This leads to the condition~\eqref{TTS-condition-AxiStationary-2}.

\item[(iii)] The case that $k_2\le k_1$.
Since $f(x)$ takes the minimum value at
one of the endpoints, we require
$f(\pm 1)\ge 0$.
This leads to the condition~\eqref{TTS-condition-AxiStationary-3}.
\end{itemize}

%
%
\section{Useful geometric formulas}
\label{appendix-c}

In Sects.~\ref{section3} and \ref{section4},
we discussed the connection between the TTS and the LTS
and the area bound of TTSs for
(i) static and (ii) stationary and 
axisymmetric cases separately.
In this appendix, we will present some useful formulas
that have a unified 
treatment for them and the potential for further generalization
to dynamical cases. 
We can also see the role of the assumption of 
staticity/stationarity in 
the argument of the main text.

Here, the setup depicted by Fig.~\ref{schematic-TTS}
is considered. Specifically, the unit normal $n^a$ to the spacelike hypersurface $\Sigma$
is tangent vectors of $S$, and the outward unit normal $\hat{r}^a$ to $S$ is tangent to $\Sigma$.
First, we derive the general formula
without assuming $t^a$ to be a Killing field or
$t^a$ to be tangent to $S$ 
(unlike the main text of this paper);
the discussion here also holds for
dynamical setups as long as the above conditions are satisfied.
After that, the static case and
the axisymmetric stationary case are considered.
We also note that the study here concerns 
the relations between geometric quantities,
and the field equations are not explicitly imposed. 
In addition to the quantities in Fig.~\ref{schematic-TTS},
we introduce one more geometric quantity, that is,
the extrinsic curvature $\bar{k}_{ab}$ of $S_0$
in the hypersurface $S$
as defined by $\bar{k}_{ab}=(1/2)\widehat{\pounds}_{n}h_{ab}$.
Here, $\widehat{\pounds}$ denotes the Lie derivative
in the timelike hypersurface $S$.

\subsection{Useful formulas for consideration on LTS}
\label{appendix-c1}

Here, we present the formulas that are useful for considering
the connection between the LTS and the TTS.
Applying the trace of the Ricci equation to $S_0$ as a hypersurface in $S$
and $\Sigma$, respectively, we have
\begin{subequations}
\begin{eqnarray}
  R_{ab}n^an^b &=& -\pounds_n K-K_{ab}K^{ab}+\frac{1}{\alpha}D^2\alpha,
  \\
R_{ab}\hat{r}^a \hat{r}^b &=& -\pounds_{\hat r} \bar K-\bar K_{ab}\bar K^{ab}-\frac{1}{\varphi}\bar D^2\varphi,
\end{eqnarray}
\end{subequations}
where $\alpha$ and $\varphi$ are the time and space lapse functions
with respect to the $n^a$ and $\hat{r}^a$ 
directions, respectively. From these two, one may want to construct 
\begin{equation}
  \pounds_{\hat r}\bar K
  =
  -\pounds_n K
  -R_{ab}(n^an^b+\hat{r}^a \hat{r}^b)
 -K_{ab}K^{ab}-\bar K_{ab}\bar K^{ab}
  +\frac{1}{\alpha}D^2\alpha-\frac{1}{\varphi}\bar D^2\varphi.
\label{liertrbarK}
\end{equation}
Recall the decomposition of $\bar{K}_{ab}$ that 
is given by Eq.~\eqref{barKab-general}
with the vector $v_a$ defined in Eq.~\eqref{definition_va_1}.
These formulas also hold  
in the current setup without assuming staticity/stationarity.
The trace of Eq.~\eqref{barKab-general} is
\begin{equation}
  \bar{K}=k+\frac{1}{\alpha}\pounds_{\hat{r}}\alpha.
  \label{barKdecomp}
\end{equation}
Substituting
these formulas together with the
decomposition of $D^2\alpha$ and $\bar{D}^2\varphi$ into
lower-dimensional quantities,
\begin{subequations}
\begin{eqnarray}
  D^2 \alpha
  &=&
  \mathcal{D}^2\alpha+k\pounds_{\hat r}\alpha
  +\frac{\mathcal{D}_a \varphi}{\varphi}\mathcal{D}^a \alpha +(\pounds_{\hat r})^2 \alpha,
  \\
  \bar{D}^2 \varphi
  &=&
  \mathcal{D}^2\varphi-\bar{k} \pounds_n \varphi
  +\mathcal{D}^a\varphi\frac{\mathcal{D}_a \alpha}{\alpha}
  -(\pounds_{n})^2 \varphi,
\end{eqnarray}
\end{subequations}
we arrive at the general formula for $\pounds_{\hat{r}}k$,
\begin{multline}
\pounds_{\hat r} k  =  -\pounds_n K
-R_{ab}(n^an^b+\hat{r}^a \hat{r}^b) 
+\frac{1}{\alpha}\mathcal{D}^2\alpha-\frac{1}{\varphi}\mathcal{D}^2\varphi
\\
+k\frac{\pounds_{\hat r} \alpha}{\alpha}
+\bar{k}\frac{\pounds_{n} \varphi}{\varphi}
+\frac{1}{\varphi}(\pounds_n)^2 \varphi
-K_{ab}K^{ab}
-k_{ab}k^{ab}
+2v_av^a.
\label{liertrk}
\end{multline}
Below, we look at 
the static case and the stationary axisymmetric
case, one by one.

\subsubsection{Static case}

Since $K_{ab}=\pounds_n \varphi=v_a=0$ for static spacetimes,
Eq.~\eqref{liertrk} becomes 
\begin{equation}
\pounds_{\hat r}k=-R_{ab}(n^an^b+{\hat r}^a {\hat r}^b)
+\frac{1}{\alpha}{\cal D}^2\alpha
-\frac{1}{\varphi}{\cal D}^2\varphi
+k \frac{{\pounds}_{\hat r}\alpha}{\alpha}-k_{ab}k^{ab}.
\end{equation}
This corresponds to Eq.~\eqref{rDtrk} when the Einstein equation holds. 

\subsubsection{Stationary and axisymmetric case}

For stationary and axisymmetric spacetimes,
adopting the time slice
on which the shift vector $\beta_a$ becomes $\beta^a=-\omega \phi^a$, 
the extrinsic curvature $K_{ab}$ of $\Sigma$ has the form
of Eq.~\eqref{KabpDo},
and $v_a$ is given as Eq.~\eqref{va_AxiStationary}. 
Since $K=\pounds_n \varphi=\phi^a D_a \omega=\hat{r}^a \phi_a=0$,
Eq.~\eqref{liertrk} becomes 
\begin{equation}
\pounds_{\hat r}k=-\frac{\phi^2}{2\alpha^2}(\mathcal{D}\omega)^2
-R_{ab}(n^an^b+\hat{r}^a \hat{r}^b)
+\frac{1}{\alpha}\mathcal{D}^2\alpha
-\frac{1}{\varphi}\mathcal{D}^2\varphi
+k \frac{\pounds_{\hat r}\alpha}{\alpha} -k_{ab}k^{ab}.
\end{equation}
This corresponds to Eq.~\eqref{rDtrk2} when the Einstein equation holds.
Note that the $t$--$\phi$ orthogonality condition
has not been used in deriving this relation.

\subsection{Useful formulas for consideration on area bound}
\label{appendix-c2}

Now, we turn our attention to the area bound.  
As introduced at the beginning of this appendix,
$\bar k_{ab}$ denotes the extrinsic curvature of $S_0$ in $S$.
From the double trace of the 
Gauss equation on $S_0$ in $S$, we have 
\begin{equation}
  {}^{(2)}R
  =
  {}^{(3)}\bar{R}
  +2 {}^{(3)}\bar{R}_{ab}n^an^b
  -\bar{k}^2
  +\bar{k}_{ab} \bar{k}^{ab}, \label{dtgauss}
\end{equation}
where ${}^{(3)} \bar{R}_{ab}$ is the Ricci tensor with respect to the metric $P_{ab}$ of $S$. Here, ${}^{(3)} \bar{R}$ is written 
using the double trace of the Gauss equation on $S$ in the spacetime,  
\begin{equation}
  {}^{(3)} \bar{R}
  =
  -2G_{ab}\hat{r}^a \hat{r}^b
  +\bar{K}^2 
  -\bar{K}_{ab} \bar{K}^{ab}, 
\end{equation}
where $G_{ab}$ is the Einstein tensor 
$G_{ab}:=R_{ab}-(1/2)g_{ab}R$. 
As for ${}^{(3)} \bar{R}_{ab} n^an^b$, one may want to employ
the following formula derived by taking trace of the Ricci equation:
\begin{equation}
{}^{(3)} \bar R_{ab} n^an^b=-\pounds_n \bar k-\bar k_{ab} \bar k^{ab}
+\frac{1}{\alpha}{\cal D}^2 \alpha.
\label{3riccrr}
\end{equation}
Here, we used the fact that the hypersurface $S$
has the same time lapse function $\alpha$ as 
that of the spacetime $\mathcal{M}$ because the timelike unit
normal $n^a$ is tangent to $S$.
Then, Eq.~\eqref{dtgauss} becomes 
\begin{equation}
{}^{(2)}R=-2 \pounds_n \bar k -2G_{ab}\hat{r}^a \hat{r}^b+\bar K^2-\bar K_{ab}\bar K^{ab}
-\bar k^2-\bar k_{ab} \bar k^{ab}+\frac{2}{\alpha}\mathcal{D}^2\alpha.
\end{equation}
Finally, we have the following formula using Eqs.~\eqref{barKab-general} and
\eqref{barKdecomp}:
\begin{equation}
  {}^{(2)}R
  =-2 \pounds_n \bar k -2G_{ab}\hat{r}^a \hat{r}^b+ k^2-k_{ab}k^{ab}
+2k\frac{\pounds_{\hat r}\alpha}{\alpha}+2v_av^a
-\bar k^2-\bar k_{ab} \bar k^{ab}+\frac{2}{\alpha}\mathcal{D}^2 \alpha.
\label{2dricci}
\end{equation}
This is the general formula for ${}^{(2)}R$.
We look at the formulas for the static case and
the stationary and axisymmetric case, one by one.

\subsubsection{Static case}

For static cases, $\bar k_{ab}=v_a=0$ holds, and 
the above formula then becomes 
\begin{equation}
{}^{(2)}R=-2G_{ab}\hat{r}^a \hat{r}^b+ k^2-k_{ab}k^{ab}+2k\frac{\pounds_{\hat r}\alpha}{\alpha}+\frac{2}{\alpha}\mathcal{D}^2 \alpha.
\end{equation}
This corresponds to Eq.~\eqref{2D-RicciScalar-static} when the Einstein equation holds. 

\subsubsection{Stationary and axisymmetric case}

For this case, since
$\bar{k}_{ab}={h_{a}}^{c}{h_{b}}^{d}K_{cd}$, we have
\begin{equation}
\bar k_{ab}=\frac{1}{2\alpha}(\phi_a {\cal D}_b \omega+\phi_b {\cal D}_a \omega)
\end{equation}
and $\bar{k}=0$. The formula for $v_a$ is given in Eq.~\eqref{va_AxiStationary}.
Using these formulas, 
Eq.~\eqref{2dricci} becomes 
\begin{equation}
{}^{(2)}R=-2G_{ab}\hat{r}^a \hat{r}^b+ k^2-k_{ab}k^{ab}
+2k\frac{\pounds_{\hat r}\alpha}{\alpha}
+\frac{2}{\alpha}\mathcal{D}^2 \alpha
+\frac{\phi^2}{2\alpha^2} [(\pounds_{\hat{r}} \omega)^2-(\mathcal{D}\omega)^2].
\end{equation}
This corresponds to Eq.~\eqref{2D-RicciScalar-AxiStationary}
when the Einstein equation holds. 
Note that the $t$--$\phi$ orthogonality property is not necessary
in order to derive this formula.


\begin{thebibliography}{9}

\bibitem{Penrose:1973} 
  R.~Penrose,
  Annals N.\ Y.\ Acad.\ Sci.\  {\bf 224}, 125 (1973).

  
\bibitem{Wald:1977} 
P.~S.~Jang and R.~M.~Wald, J.\ Math.\ Phys.\ {\bf 18}, 41 (1977). 



\bibitem{Huisken:2001} 
G.~Huisken and T.~Ilmanen, J.\ Diff.\ Geom.\ {\bf 59}, 353 (2001). 

  
\bibitem{Bray:2001} 
H.~Bray, J.\ Diff.\ Geom.\ {\bf 59}, 177 (2001).

  
\bibitem{Virbhadra:1999} 
  K.~S.~Virbhadra and G.~F.~R.~Ellis,
  Phys.\ Rev.\ D {\bf 62}, 084003 (2000)
  [arXiv:astro-ph/9904193].


  
\bibitem{Cardoso:2016} 
  V.~Cardoso, E.~Franzin and P.~Pani,
  Phys.\ Rev.\ Lett.\  {\bf 116}, 171101 (2016);
  Phys.\ Rev.\ Lett.\  {\bf 117}, 089902 (2016) [erratum]
  [arXiv:1602.07309 [gr-qc]].


  

\bibitem{Synge:1966}
J.~L.~Synge, Mon.\ Not.\ Roy.\ Soc.\ {\bf 131}, 463 (1966). 

\bibitem{Perlick:2004}
  V.~Perlick,
  Living Rev. Relativity {\bf 7},  9 (2004).


  
\bibitem{Claudel:2000} 
  C.~M.~Claudel, K.~S.~Virbhadra and G.~F.~R.~Ellis,
  J.\ Math.\ Phys.\  {\bf 42}, 818 (2001)
  [arXiv:gr-qc/0005050].

  

\bibitem{Cederbaum:2014} 
  C.~Cederbaum,
  arXiv:1406.5475 [math.DG].
  
\bibitem{Cederbaum:2015a} 
  C.~Cederbaum and G.~J.~Galloway,
  arXiv:1504.05804 [math.DG].

\bibitem{Yazadjiev:2015a} 
  S.~Yazadjiev and B.~Lazov,
  Classical\ Quantum\ Gravity\  {\bf 32}, 165021 (2015)
  [arXiv:1503.06828 [gr-qc]].

  
  \bibitem{Cederbaum:2015b} 
  C.~Cederbaum and G.~J.~Galloway,
  Classical\ Quantum\ Gravity\  {\bf 33}, 075006 (2016)
  [arXiv:1508.00355 [math.DG]].

\bibitem{Yazadjiev:2015b} 
  S.~S.~Yazadjiev,
  Phys.\ Rev.\ D {\bf 91}, 
  123013 (2015)
  [arXiv:1501.06837 [gr-qc]].

\bibitem{Yazadjiev:2015c} 
  S.~Yazadjiev and B.~Lazov,
  Phys.\ Rev.\ D {\bf 93}, 
  083002 (2016)
  [arXiv:1510.04022 [gr-qc]].

\bibitem{Rogatko:2016} 
  M.~Rogatko,
  Phys.\ Rev.\ D {\bf 93}, 
  064003 (2016)
  [arXiv:1602.03270 [hep-th]].

\bibitem{Yoshino:2016}
  H.~Yoshino,
  Phys.\ Rev.\ D {\bf 95}, 
  044047 (2017)
  [arXiv:1607.07133 [gr-qc]].

\bibitem{Tomikawa:2016} 
  Y.~Tomikawa, T.~Shiromizu, and K.~Izumi,
   Prog.\ Theor.\ Exp.\ Phys.\ {\bf 2017}, 033E03 (2017)
  [arXiv:1612.01228 [gr-qc]].

  
\bibitem{Tomikawa:2017} 
  Y.~Tomikawa, T.~Shiromizu and K.~Izumi,
  arXiv:1702.05682 [gr-qc].
  
\bibitem{Gibbons:2016} 
  G.~W.~Gibbons and C.~M.~Warnick,
  Phys.\ Lett.\ B {\bf 763}, 169 (2016)
  [arXiv:1609.01673 [gr-qc]].
  
\bibitem{Teo:2003} 
  E.~Teo,
  Gen.\ Relativ.\ Gravit.\  {\bf 35}, 1909 (2003).

\bibitem{Shiromizu:2017}
  T.~Shiromizu, Y.~Tomikawa, K.~Izumi, and H.~Yoshino,
  Prog.\ Theor.\ Exp.\ Phys.\ {\bf 2017}, 033E01 (2017)
  [arXiv:1701.00564 [gr-qc]].

  
\bibitem{Geroch:1973} 
R.~Geroch, Ann.\ N.Y.\ Acad.\ Sci.\ {\bf 224}, 108 (1973). 

\bibitem{Wang:2001}
  X.~Wang,  J.\ Diff.\ Geom.\ {\bf 57}, 273 (2001). 

  
\bibitem{Wald}
 R.~Wald, {\it General Relativity}
 (The University of Chicago Press, Chicago, 1984).

 
\bibitem{Gullstrand:1922}
  A.~Gullstrand,
  Ark.\ Mat.\ Astron.\ Fys {\bf 16}, 1 (1922). 

  
\bibitem{Carter:1970} 
  B.~Carter,
  Commun.\ Math.\ Phys.\  {\bf 17}, 233 (1970).

  
\bibitem{Carter:1969} 
  B.~Carter,
  J.\ Math.\ Phys.\  {\bf 10}, 70 (1969).
  
\bibitem{Kerr:1963} 
  R.~P.~Kerr,
  Phys.\ Rev.\ Lett.\  {\bf 11}, 237 (1963).


  
\bibitem{Doran:1999} 
  C.~Doran,
  Phys.\ Rev.\ D {\bf 61}, 067503 (2000)
  [arXiv:gr-qc/9910099].

  

\bibitem{Frolov:1998}
  V.~P.~Frolov and I.~D.~Novikov,
  {\it Black Hole Physics: Basic Concepts and New Developments} 
  (Kluwer Academic Publishers, Dordrecht and Boston, 1998).

  
\bibitem{Ansorg:2005} 
  M.~Ansorg and D.~Petroff,
  Phys.\ Rev.\ D {\bf 72}, 024019 (2005)
  [arXiv:gr-qc/0505060].

  
\bibitem{Herdeiro:2014} 
  C.~A.~R.~Herdeiro and E.~Radu,
  Phys.\ Rev.\ Lett.\  {\bf 112}, 221101 (2014)
  [arXiv:1403.2757 [gr-qc]].

\bibitem{Herdeiro:2015} 
  C.~Herdeiro and E.~Radu,
  Class.\ Quant.\ Grav.\  {\bf 32}, 144001 (2015)
  [arXiv:1501.04319 [gr-qc]].
  
  
\bibitem{Herdeiro:2016} 
  C.~Herdeiro, E.~Radu and H.~Runarsson,
  Class.\ Quant.\ Grav.\  {\bf 33}, 
  154001 (2016)
  [arXiv:1603.02687 [gr-qc]].

  
\bibitem{Horvat:2013} 
  D.~Horvat, S.~Ilijic, A.~Kirin, and Z.~Narancic,
  Class.\ Quant.\ Grav.\  {\bf 30}, 095014 (2013)
  [arXiv:1302.4369 [gr-qc]].

  
  
\bibitem{Virbhadra:2002} 
  K.~S.~Virbhadra and G.~F.~R.~Ellis,
  Phys.\ Rev.\ D {\bf 65}, 103004 (2002).
  
  
\bibitem{Bozza:2002} 
  V.~Bozza,
  Phys.\ Rev.\ D {\bf 66}, 103001 (2002)
  [arXiv:gr-qc/0208075].

  
\bibitem{Virbhadra:2007} 
  K.~S.~Virbhadra and C.~R.~Keeton,
  Phys.\ Rev.\ D {\bf 77}, 124014 (2008)
  [arXiv:0710.2333 [gr-qc]].

  
\bibitem{Sahu:2012} 
  S.~Sahu, M.~Patil, D.~Narasimha, and P.~S.~Joshi,
  Phys.\ Rev.\ D {\bf 86}, 063010 (2012)
  [arXiv:1206.3077 [gr-qc]].

  
\bibitem{Sahu:2013} 
  S.~Sahu, M.~Patil, D.~Narasimha, and P.~S.~Joshi,
  Phys.\ Rev.\ D {\bf 88}, 103002 (2013)
  [arXiv:1310.5350 [gr-qc]].

  

\bibitem{Dain:2002} 
  S.~Dain, C.~O.~Lousto, and R.~Takahashi,
  Phys.\ Rev.\ D {\bf 65}, 104038 (2002)
  [arXiv:gr-qc/0201062].
  

\bibitem{Dain:2011} 
  S.~Dain and M.~Reiris,
  Phys.\ Rev.\ Lett.\  {\bf 107}, 051101 (2011)
  [arXiv:1102.5215 [gr-qc]].

  
  
\bibitem{Thorne:1972}K.~S.~Thorne, ~in {\it Magic without Magic: John
    Archbald Wheeler}, ed. J.Klauder (Freeman,~San Francisco,
  1972).

  
\bibitem{Yoshino:2007}
  H.~Yoshino,
  Phys.\ Rev.\ D {\bf 77}, 041501 (2008) 
  [arXiv:0712.3907 [gr-qc]].


\bibitem{Saida:2015}
  H.~Saida, A.~Fujisawa, C.~M.~Yoo, and Y.~Nambu,
  Prog.\ Theor.\ Exp.\ Phys.\ {\bf 2016},  
  043E02 (2016)
  [arXiv:1503.01840 [gr-qc]].
  

\end{thebibliography}
\end{document}